\documentclass[english,aps,pre,twocolumn]{revtex4}
\usepackage[T1]{fontenc}
\usepackage[latin9]{inputenc}
\usepackage{graphicx}
\usepackage{amssymb}

\makeatletter

\providecommand{\tabularnewline}{\\}

\@ifundefined{textcolor}{}
{%
 \definecolor{BLACK}{gray}{0}
 \definecolor{WHITE}{gray}{1}
 \definecolor{RED}{rgb}{1,0,0}
 \definecolor{GREEN}{rgb}{0,1,0}
 \definecolor{BLUE}{rgb}{0,0,1}
 \definecolor{CYAN}{cmyk}{1,0,0,0}
 \definecolor{MAGENTA}{cmyk}{0,1,0,0}
 \definecolor{YELLOW}{cmyk}{0,0,1,0}
 }

\makeatother

\usepackage{babel}

\begin{document}

\title{Numerical Results for Spin Glass Ground States on Bethe Lattices:
Gaussian Bonds}

\author{Stefan Boettcher\footnote{http://www.physics.emory.edu/faculty/boettcher/}}

\affiliation{Physics Department, Emory University, Atlanta, Georgia 30322; USA}

\begin{abstract}
The average ground state energies for spin glasses on Bethe lattices
of connectivities $r=3,\ldots,15$ are studied numerically for a
Gaussian bond distribution. The Extremal Optimization heuristic is
employed which provides high-quality approximations to ground states.
The energies obtained from extrapolation to the thermodynamic limit
smoothly approach the ground-state energy of the
Sherrington-Kirkpatrick model for $r\to\infty$. Consistently for all
values of $r$ in this study, finite-size corrections are found to
decay approximately with $\sim N^{-4/5}$.  The possibility of $\sim
N^{-2/3}$ corrections, found previously for Bethe lattices with a
bimodal $\pm J$ bond distribution and also for the
Sherrington-Kirkpatrick model, are constrained to the additional
assumption of very specific higher-order terms.  Instance-to-instance
fluctuations in the ground state energy appear to be asymmetric up to
the limit of the accuracy of our heuristic.  The data analysis
provides insights into the origin of trivial fluctuations when using
continuous bonds and/or sparse networks.
\end{abstract}


\maketitle

\section{Introduction\label{sec:Introduction}}

We study the ground state ($T=0$) properties of spin glasses on Bethe
lattices with Gaussian bond distribution, which are also considered at
low but finite temperatures in Ref.~\cite{Rizzo09b}. In many ways,
this study resembles that for the bimodal $\pm J$ bond distribution in
Refs.~\cite{Boettcher03a,Boettcher03b}. Yet, surprisingly, the
behavior for finite-size corrections differ significantly between both
distributions.

Bethe lattices are $r$-regular graphs~\cite{Bollobas}, i.~e. randomly
connected graphs consisting of $N$ vertices, each having a fixed
number, $r$, of neighbors~\cite{Mezard03}.  We explore the large-$N$
regime of low-connectivity graphs, $r=3,\ldots,15$, which are of great
theoretical interest as finite-connected, mean-field models for
low-dimensional lattice spin glasses~\cite{MPV,Viana85}.  A great
number of studies have focused on various aspects of this conceptually
simple model to hone the complex mathematical techniques required to
treat disordered
systems~\cite{Mezard87,mezard:01,Mezard03,Tria02,dedominicis:89,Mottishaw87,Lai90}
or optimization
problems~\cite{Monasson99,MPZ,Franz01,Wong87,Banavar87b,Zdeborova10}.

As before in Refs.~\cite{Boettcher03a,Boettcher03b}, we use the
Extremal Optimization (EO)
heuristic~\cite{Boettcher00,Boettcher01a,Dagstuhl04} to find
approximations to spin glass ground states. In previous papers, we
have demonstrated the capabilities of EO in determining near-optimal
solutions for spin glasses, the coloring
problem~\cite{Boettcher01a,Boettcher04a} and the graph partitioning
problem~\cite{Boettcher99a,Boettcher00,Boettcher01b,Percus08,Zdeborova10}.
It is generally harder to find good approximations in complex energy
landscapes with a local search heuristic, such as EO, for a problem
with continuous weights~\cite{Bauke04}, but some encouraging results
exist~\cite{Middleton04}. While our results appear to be sufficiently
accurate for the prediction of energy averages for systems up to size
$N=2048$, more detailed features, such as their fluctuations over the
ensemble of instances (requiring higher moments of the energy), are
less reliable for system sizes $N>256$ and larger degree $r$. Hence,
for Ising spin glass simulations with EO, discrete $\pm J$ bond
weights are usually preferable. Based on the experience with the
Sherrington-Kirkpatrick (SK) model, scaling properties of
thermodynamic observables are generally believed to be universal,
independent of the details of the bond distribution used. There, for
instance, finite-size corrections are found to scale approximately
with $\sim
N^{-2/3}$~\cite{EOSK,Bouchaud03,Katzgraber05,Palassini08,Aspelmeier07}
(just as for the Bethe lattices with bimodal bonds) and even the
average ground-state energy density $\left\langle e_{SK}\right\rangle
\approx-0.76317\ldots$ is universal for any symmetric bond
distribution~\cite{MPV}. It would be therefore remarkable to find such
distinct scaling behavior between distributions at the level of
finite-size corrections, as this investigation suggest. Some previous
investigations of Bethe lattices with Gaussian
bonds~\cite{Bouchaud03,Liers03} have been consistent with
$N^{-2/3}$-corrections but were based on smaller sizes and
significantly less statistics as in our study here.  As a possible
resolution, we would have to appeal to ad-hoc assumptions about
higher-order corrections.  (The distinct effects of discrete versus
continuous bonds on defect energies in finite-dimensional lattice spin
glasses have been studied numerically in Ref.~\cite{Hartmann01}, and
also with the renomalization group in Ref.~\cite{amoruso:03}.) In
turn, the variation of corrections with the details of the bond
distribution may provide important clues towards extending replica
theory to include finite-size effects.

While the value of this work lies in exploring the range of
finite-size scaling on sparse random graphs for spin glasses with
differing distributions, it also provides a cautionary note about the
analysis of data obtained from such systems. In a sparse system, such
as a random graph or a randomly diluted lattice, (trivial) normal
fluctuations in geometry \footnote{The fluctuations referred to are
  those in the total number of bonds in a graph over the ensemble of
  graphs. These are normal, even if the (independent) degrees at each
  vertex follows a separate distribution of non-zero width, such as a
  Poissonian.} or in the bond distribution can obscure the physical
essence of the problem at hand. We will argue that in such a system we
need to focus attention to the actual {}``cost'' $C$ of disorder in
terms of the frustration, instead of the energy $E$ itself. Just
consider a simple ferromagnet with fixed bonds $J$ on an ordinary
random graph $G_{N,p}$~\cite{Bollobas}, or alternatively, with
continuously distributed bonds $J>0$ on a fixed-degree Bethe lattice
of size $N$. In either case, the ground state energy has all bonds
satisfied, i.~e. no cost in the number of frustrated bonds ($C=0$),
and ground state cost fluctuations exhibit a $\delta$-peak,
correspondingly. But by the central limit theorem, the absolute sum of
all bonds
\begin{equation}
B=\sum_{i=1}^{rN/2}\left|J_{i}\right|
\label{Beq}
\end{equation}
has a normal distribution, inherited by the ground state energy
fluctuations via $E=2C-B$. We would claim that the relevant
fluctuations at finite $N$ are captured by $C$, not $E$, although the
thermodynamic averages $\langle E\rangle$ and $\langle C\rangle$ are
completely equivalent, of course. In any case, careful distinction is
advisable in an environment of competing finite-size corrections,
especially when averaging inherently finite samples in simulations.

The effect of these trivial fluctuations is most pronounced in the
study of fluctuations in the ground states of the systems, which have
been intensely studied in recent
years~\cite{Bouchaud03,andreanov:04,EOSK,Boettcher05e,Katzgraber05,Aspelmeier07,Palassini08,Parisi08,Parisi09,Rizzo09,Rizzo09b}.
Our numerical data here shows that those fluctuations in the
\emph{energy} density would predict a seemingly interesting cross-over
between non-trivial to trivial scaling between system size and
degree. In contrast, the cost density fluctuations would predict a
slow drift towards triviality at larger system sizes that is setting
in earlier for larger system sizes and might be attributable to a
decay in accuracy.  As these results are somewhat inconclusive, in a
related publication~\cite{Bo_unpub} we will show that these
fluctuations on Bethe lattices with a \emph{bimodal} distribution of
bonds behave, again, similar to those of the SK model.

In the following Sec.~\ref{sec:Spin-Glasses-on}, we introduce first
the Bethe lattices we used in the numerical calculations. Then, we
address the relation between energies and costs in
Sec.~\ref{sec:Sampling-Ground-States}.  In
Sec.~\ref{sec:-EO-Algorithm}, we briefly describe the EO algorithm,
which is amply discussed
elsewhere~\cite{Boettcher00,Boettcher01a,Dagstuhl04}.  In
Sec.~\ref{sec:Numerical-Results}, we finally present our numerical
results. Some conclusions are presented in Sec.~\ref{sec:Conclusion}.

\section{Spin Glasses on Bethe Lattices\label{sec:Spin-Glasses-on}}

Disordered spin systems on random graphs have been investigated as
mean-field models for low-dimensional spin glasses or optimization
problems, since variables are long-range connected yet have a small
number of neighbors. Particularly simple are Bethe lattices of fixed
vertex degree $r=k+1$~\cite{Mezard87,mezard:01,Mezard03}, which are
locally tree-like with vertices imagined as possessing one
up-direction and $k$ downward branches. Yet, all $r$ directions are
fully equivalent in Bethe lattices, and there is no root vertex or any
boundary. In comparison to the otherwise
more familiar random graphs studied by Erdös and
Rény~\cite{Bollobas}, Bethe lattices at a given $N$ and $r$ avoid
fluctuations in the vertex degree and in the total number of bonds.

There are slight variations in the generation of Bethe lattices. For
instance, to add a bond one could choose at random two vertices of
connectivities $<r$ to link until all vertices are $r$-connected.
Instead, we have used the method described in Ref.~\cite{Bollobas} to
generate these graphs. Here, all the terminals on the vertices form a
list of $rN$ independent variables. For each added bond, two
available terminals are chosen at random to be linked and removed from
the list. Furthermore, for algorithmic convenience, we reject graphs
which possess self-loops, i.~e. bonds that connect two terminals of
the same vertex. Multiple bonds between any pair of vertices are
allowed; otherwise it is too hard to generate feasible graphs for
small $N$, especially at larger $r$. Since $r$ remains finite for
$N\to\infty$, the energy and entropy per spin would only be effected
to $O(1/N)$ by  differences between these choices.

\section{Sampling Ground States on a sparse Graph\label{sec:Sampling-Ground-States}}

\begin{figure}
\includegraphics[clip,scale=0.35]{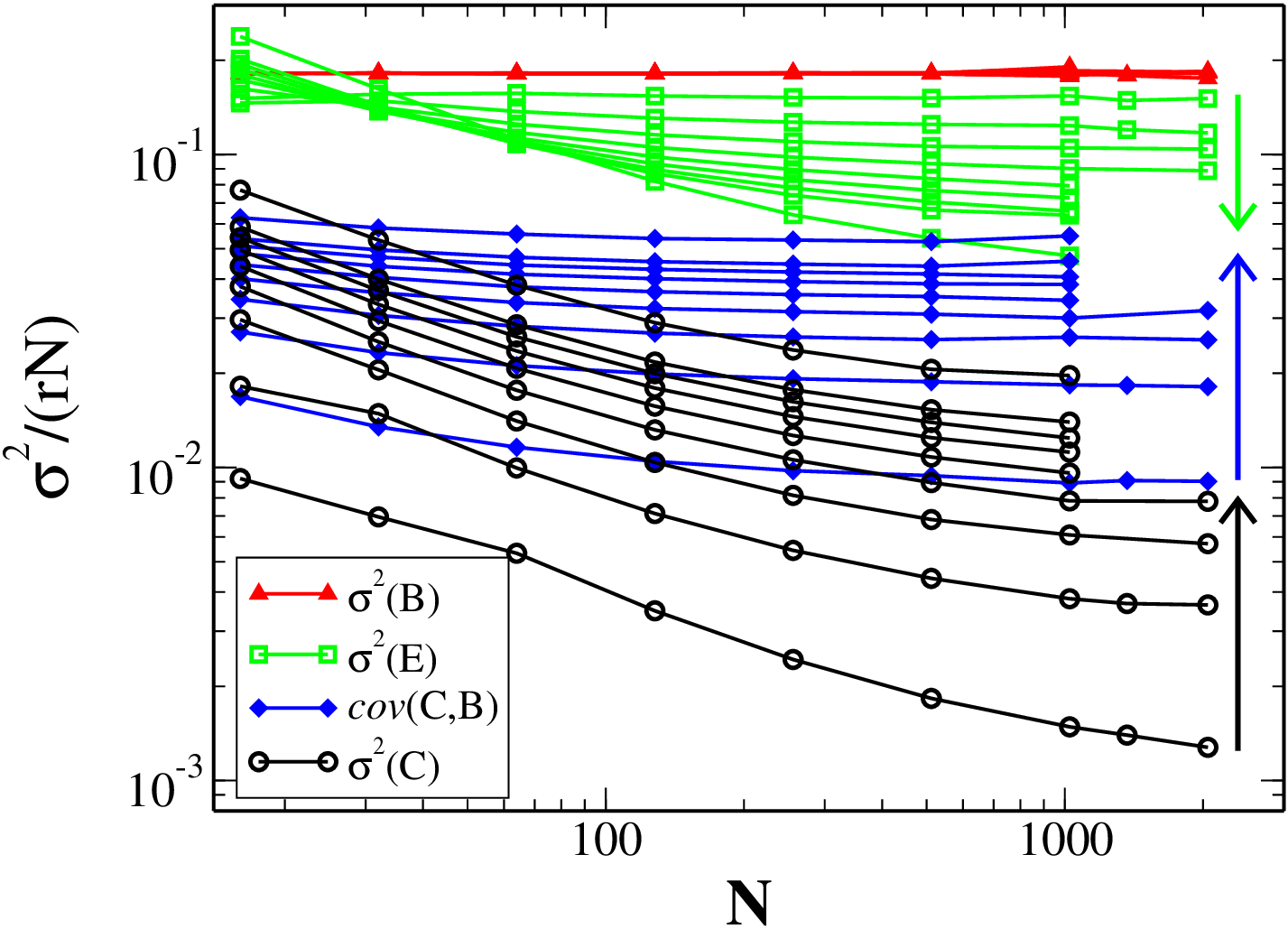}
\caption{\label{fig:covar_test}
Plot of the variances and co-variances, appropriately rescaled, in
Eq.~(\ref{eq:varianceE}) as a function of $N$ for
$r=3,4,\ldots,10,15$. (Arrows indicate increasing $r$-values for the
adjacent data sets.)  All curves are either constant in $N$
throughout, or appear to approach a constant. In this scaling, the
variance for the bond weights $B$ (red $\triangle$) collapses for all
$r$ to
$\sigma^{2}(B)/(rN)=\sigma^{2}\left(\left|J\right|\right)/2=\frac{1}{2}-\frac{1}{\pi}$,
and the variances for the energies (green $\square$) are of similar
magnitude but declining for larger $r$ (top to bottom). At some fixed
(larger) $N$, both, the covariance between $B$ and $C$ (blue
$\diamond$) as well as the variance of $C$ (black $\circ$) alone, are
quite small, but they are \emph{increasing} with $r$ (bottom to top).}
\end{figure}

Once a graphical instance is generated, be it a Bethe lattice or any
other sparse graph, we assign bonds $J_{i,j}$, here randomly chosen
from a Gaussian distribution of zero mean and unit variance, to
existing links between neighboring vertices $i$ and $j$. Each vertex
$i$ is occupied by an Ising spin variable $x_{i}\in\{-1,+1\}$. The
Hamiltonian
\begin{eqnarray}
H=-\sum_{\{bonds\}}J_{i,j}x_{i}x_{j}.
\label{Heq}
\end{eqnarray}
provides the energy of the system. For each instance $I$, the energy
$E^{(I)}$ is defined as the difference in the absolute weight of
all violated bonds, $C^{(I)}$ (the {}``cost''), and satisfied bonds,
$S^{(I)}$, i.~e. 
\begin{eqnarray}
E^{(I)}=C^{(I)}-S^{(I)};
\label{Eeq}
\end{eqnarray}
the larger the satisfied bond-weight $S^{(I)}$ in the instance, the
lower its energy $E^{(I)}$. While $C^{(I)}$ and $S^{(I)}$
vary depending on the spin configuration, for each instance
\begin{eqnarray}
B^{(I)} & = & C^{(I)}+S^{(I)}
\label{eq:BI}
\end{eqnarray}
is a constant, with $B^{(I)}$  given by Eq.~(\ref{Beq}).
Hence,
\begin{eqnarray}
E^{(I)}=2C^{(I)}-B^{(I)},
\label{E2eq}
\end{eqnarray}
provides a direct relation between cost and energy of each instance.
Thus, after averaging over a sample $n_{I}$ of instances $I$ of
size $N$ (denoted by $\langle\ldots\rangle_{N}$), we obtain by definition
for the averages
\begin{eqnarray}
\left\langle E\right\rangle _{N} & = & 2\left\langle C\right\rangle
_{N}-\left\langle B\right\rangle _{N}
\label{eq:averageE}
\end{eqnarray}
and for the variances:
\begin{eqnarray}
\sigma_{N}^{2}\left(E\right) & = &
4\sigma_{N}^{2}\left(C\right)+\sigma_{N}^{2}\left(B\right)-4cov_{N}\left(C,B\right).
\label{eq:varianceE}
\end{eqnarray}
We observe that for the sparse-graph systems with continuous bond
weights under consideration here {[}similar to the ferromagnetic example
in the Introduction, where
$\sigma_{N}^{2}\left(B\right)=\sigma_{N}^{2}\left(E\right)\sim N$ and $\sigma_{N}^{2}\left(C\right)=cov_{N}\left(C,B\right)\equiv0${]},
\begin{eqnarray}
\sigma_{N}^{2}\left(C\right) & \lesssim &
cov_{N}\left(C,B\right)\ll\sigma_{N}^{2}\left(B\right),
\label{eq:smallCOV}
\end{eqnarray}
which we demonstrate in Fig.~\ref{fig:covar_test}. Note that this
condition becomes less satisfied when graphs get denser, i.~e. $r$
increases. That trend is indicated by arrows in
Fig.~\ref{fig:covar_test}. The situation described by
Eq.~(\ref{eq:smallCOV}) does not impact the relation between the
averages in Eq.~(\ref{eq:averageE}) in any way, in particular, not the
thermodynamic quantities and their finite-size corrections discussed
below. But it does significantly affect both, the error analysis for
that data and the interpretation of the ground state energy
fluctuations, each derived from the variance.  Similarly,
$\sigma_{N}^{2}\left(B\right)$ can not be neglected in a fluctuating
geometry, such as a random graph or a random-diluted
lattice~\cite{Boettcher04c,Boettcher04b}, even if the individual
bond-weights are sharp, $\left|J_{i,j}\right|\equiv J_{0}$ with
constant $J_{0}$.

Eq.~(\ref{eq:smallCOV}) does not apply for a discrete bond
distribution on a Bethe lattice~\cite{Boettcher03a,Boettcher03b},
where $\sigma_{N}^{2}\left(B\right)\equiv0$; $E$ and $C$ are then
equivalent stochastic variables in every respect.  Furthermore, in a
dense graph such as the SK model~\cite{Sherrington75}, and independent
of any symmetric bond distribution, this situation is entirely
distinct, as even for ground states almost exactly half of all bonds
are violated, i.~e. $2C^{(I)}\approx B^{(I)}$ for large $N$, and the
energy appears merely as an extreme value within the normal
fluctuations of $2C$ and $B$. In this case, using
$C$ to study fluctuations would be futile. As $r$ is increasing, this
trend is already visible in Fig.~\ref{fig:covar_test}: the gap between
the variance in $E$ and $C$ is closing, and is bound to cross over at some
larger degree $r$.

Eqs.~(\ref{eq:varianceE}-\ref{eq:smallCOV}) imply
that the standard error of the average energy is  dominated by the error in $B$,
\begin{eqnarray}
\Delta E_{N} & = &
\frac{\sigma_{N}\left(E\right)}{\sqrt{n_{I}}}\sim\Delta
B\left(1-\frac{2cov_{N}\left(C,B\right)}{\sigma_{N}^{2}\left(B\right)}\right),
\label{eq:DeltaE}
\end{eqnarray}
which by Eq.~(\ref{eq:smallCOV}) is larger than that of $C$ for the
range of degrees $r$ used in this study. Therefore, our strategy for
determining the average ground state energy in the thermodynamic limit
will be based on an extrapolation for large $N$ of the values for
$\left\langle C\right\rangle _{N}$ and the evaluation of
Eq.~(\ref{eq:averageE}) at $N=\infty$ using the exact value of the
bond-density $\left\langle b\right\rangle _{\infty}$, here,
\begin{eqnarray}
\left\langle b\right\rangle _{\infty} &
=\lim_{N\to\infty}\frac{\left\langle B\right\rangle _{N}}{N}= &
\frac{r}{2}\left\langle \left|J\right|\right\rangle
_{\infty}=\frac{r}{\sqrt{2\pi}}\qquad
\label{eq:averageB}
\end{eqnarray}
for Bethe lattices with a Gaussian bond distribution.

More significantly, when Eq.~(\ref{eq:smallCOV}) holds, the variance
in the ground state energy fluctuations
tracks the variance of $B$, making it apparently trivial (i.~e. normal):
\begin{eqnarray}
\sigma_{N}^{2}\left(E\right) & \sim & \sigma_{N}^{2}\left(B\right)\sim N.
\label{eq:EvarN}
\end{eqnarray}
The scaling of $\sigma_{N}\left(E\right)\sim N^{1-\rho}$ has been the
focus of keen interest for various mean-field models
recently~\cite{Bouchaud03,andreanov:04,EOSK,Boettcher05e,Katzgraber05,Aspelmeier07,Palassini08,Parisi08,Parisi09,Rizzo09,Rizzo09b}. There,
non-trivial behavior is typically associated with an exponent that
obeys $\rho>\frac{1}{2}$.) We claim that due to Eq.~(\ref{eq:EvarN})
any non-trivial deviation for sparse graphs would have to be found in
the cost $C$, if it exists at all.

In the following, we will therefore focus on the
cost per spin $c=C/N$. We will study the finite-size corrections of the form
\begin{eqnarray}
\langle c\rangle_{N} & =\frac{\langle C\rangle_{N}}{N} & \sim a+\frac{b}{N^{\omega}}\left[1+\epsilon(N)\right]
\label{eq:FSS}
\end{eqnarray}
with $a\approx\langle c\rangle_{\infty}$, even taking some
higher-order corrections $\epsilon(N)\ll1$ into account.  Additionally,
we consider the fluctuations in the ground-state \emph{cost} density,
in particular, the scaling of its deviation with finite size,
\begin{eqnarray}
\sigma_{N}\left(c\right) & \sim & N^{-\rho}.
\label{eq:rho}
\end{eqnarray}

\begin{figure*}
\includegraphics[clip,scale=0.25]{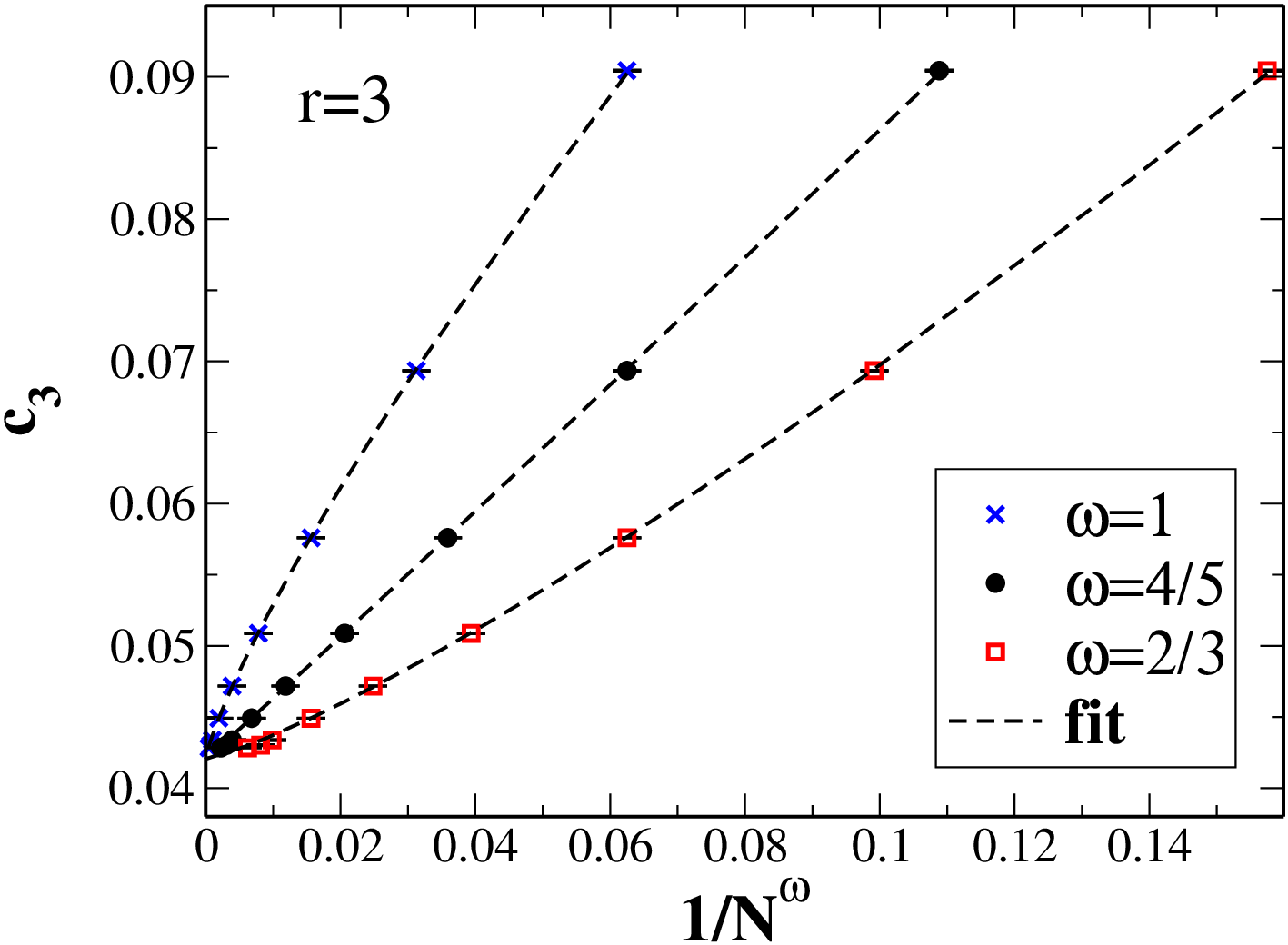}\includegraphics[clip,scale=0.25]{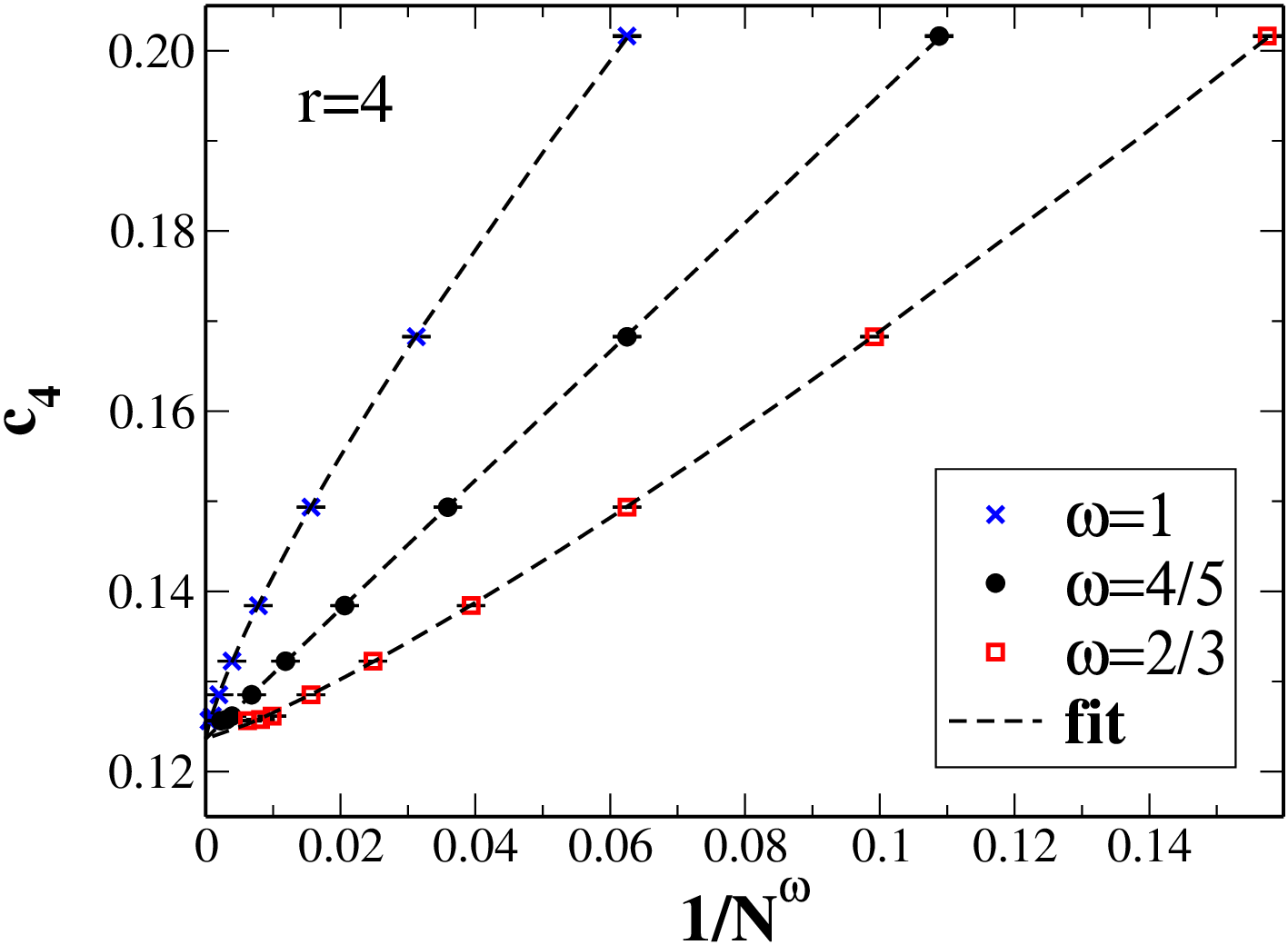}\includegraphics[clip,scale=0.25]{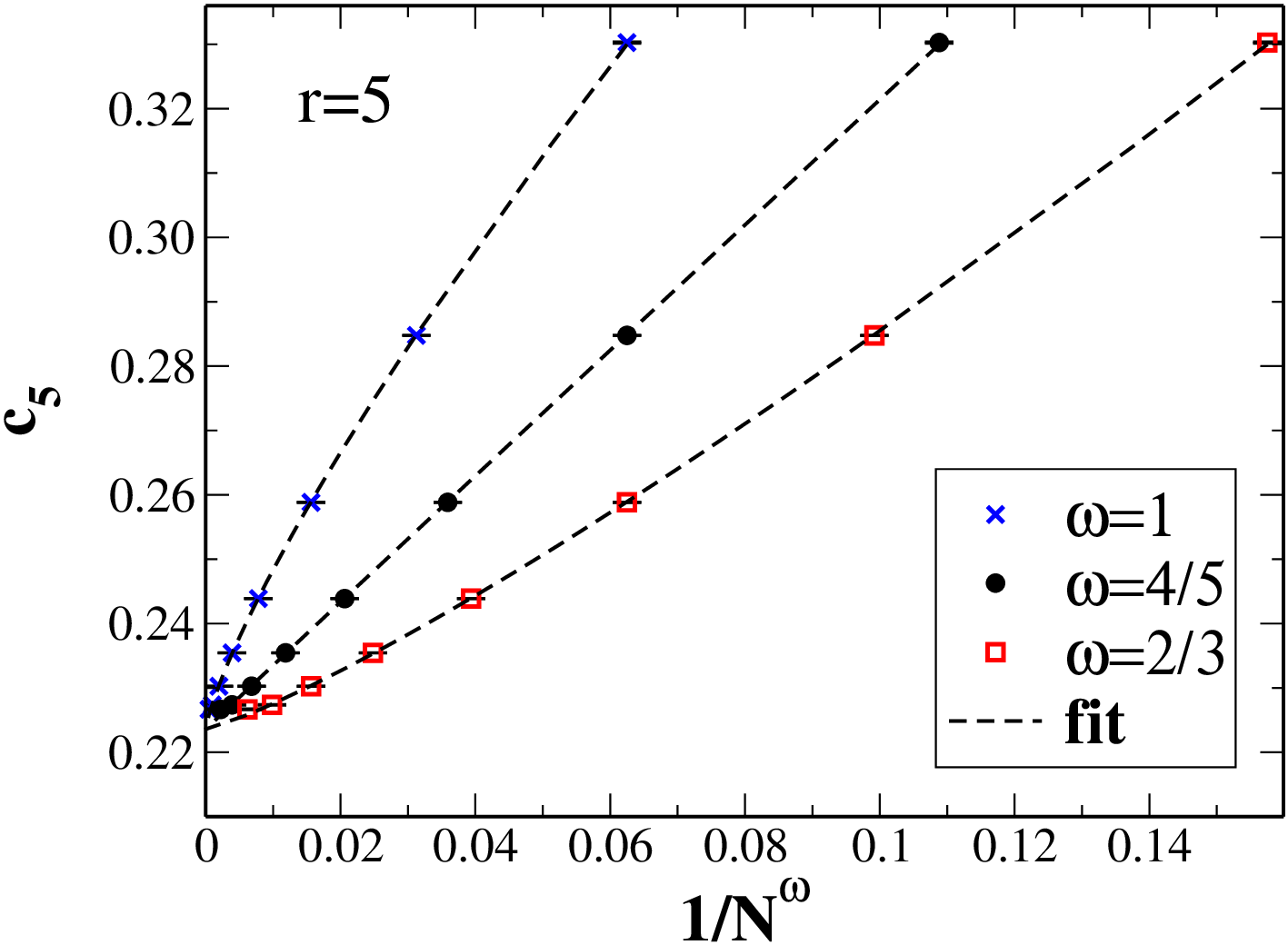}

\includegraphics[clip,scale=0.25]{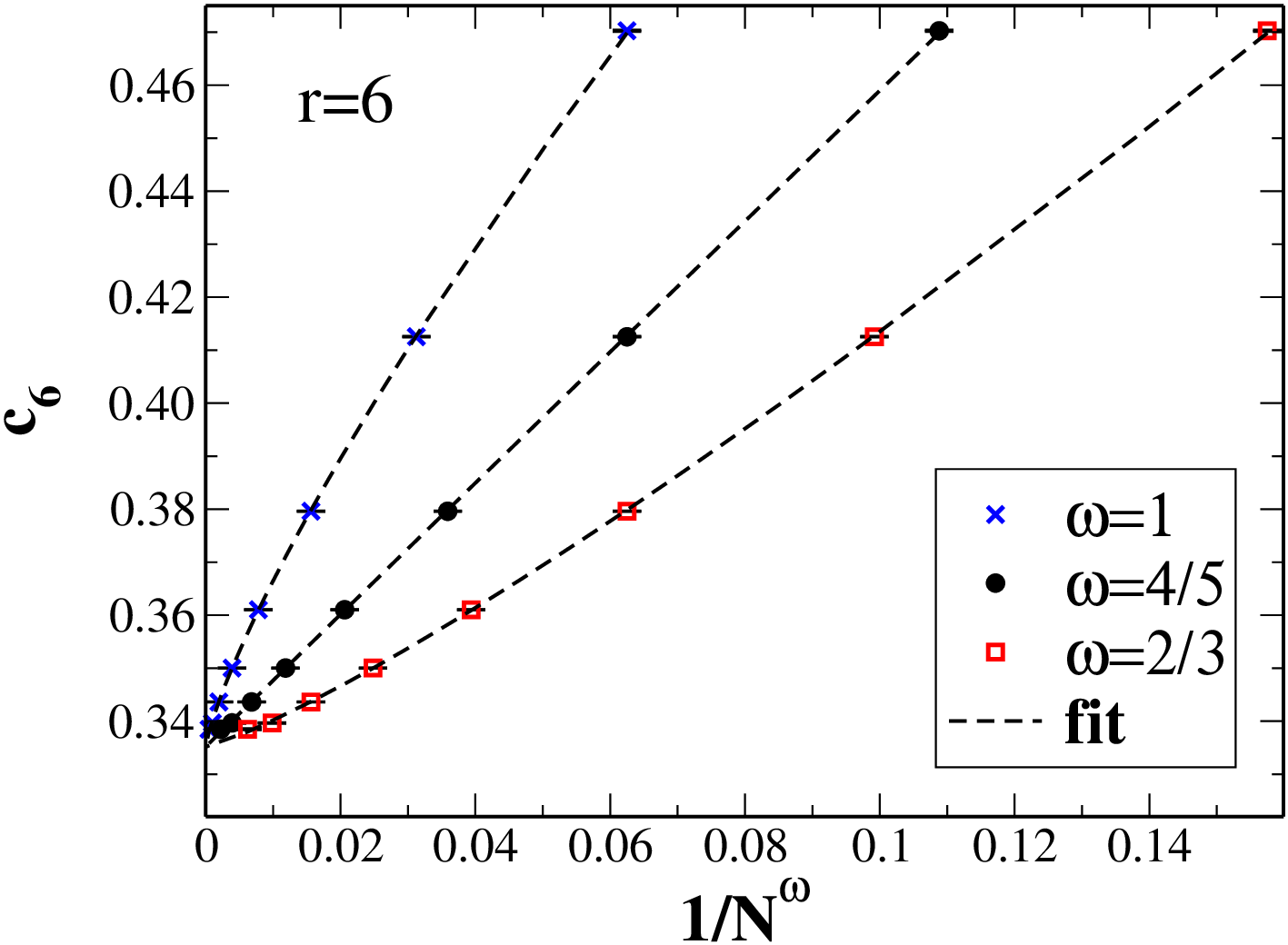}\includegraphics[clip,scale=0.25]{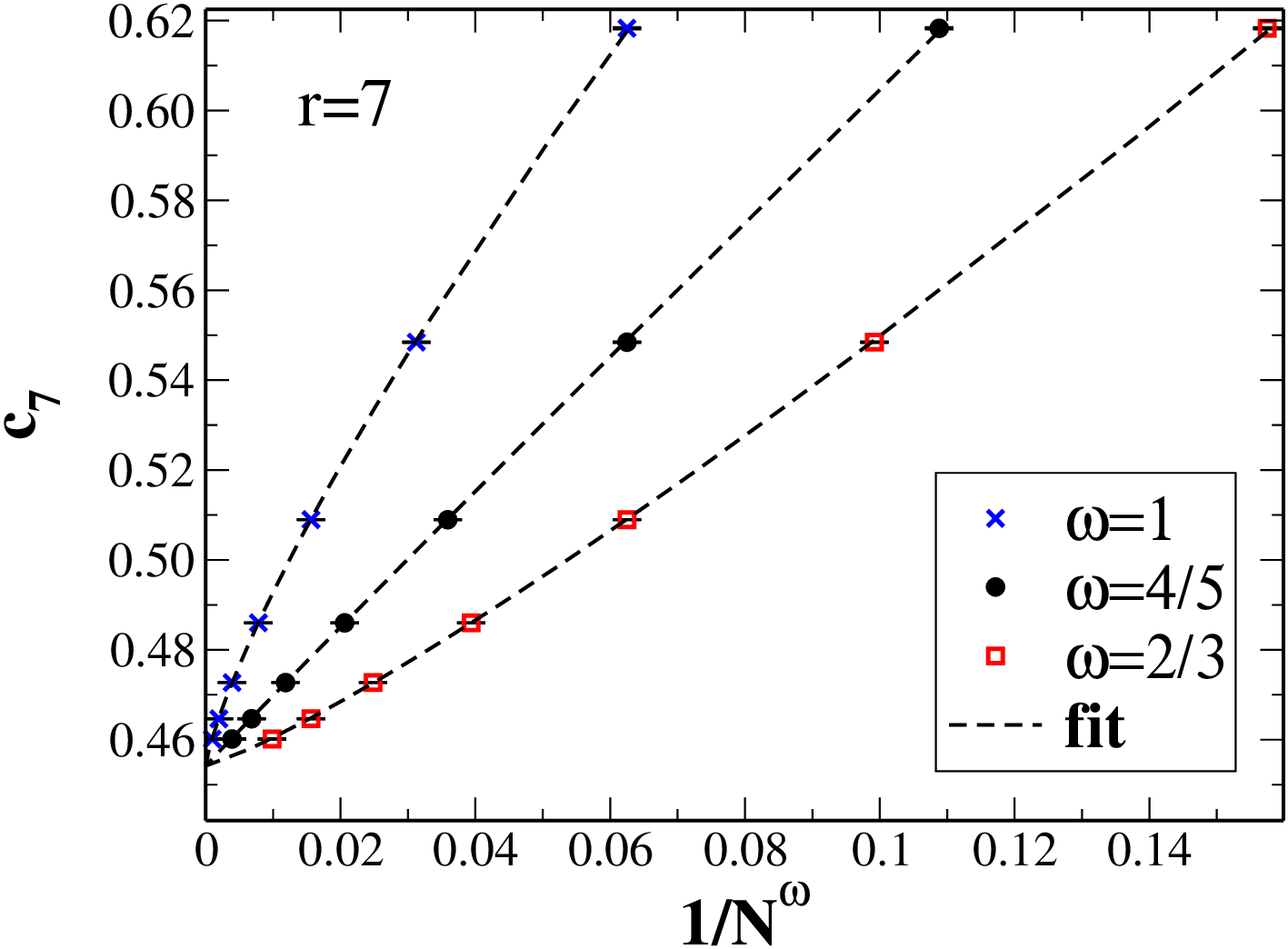}\includegraphics[clip,scale=0.25]{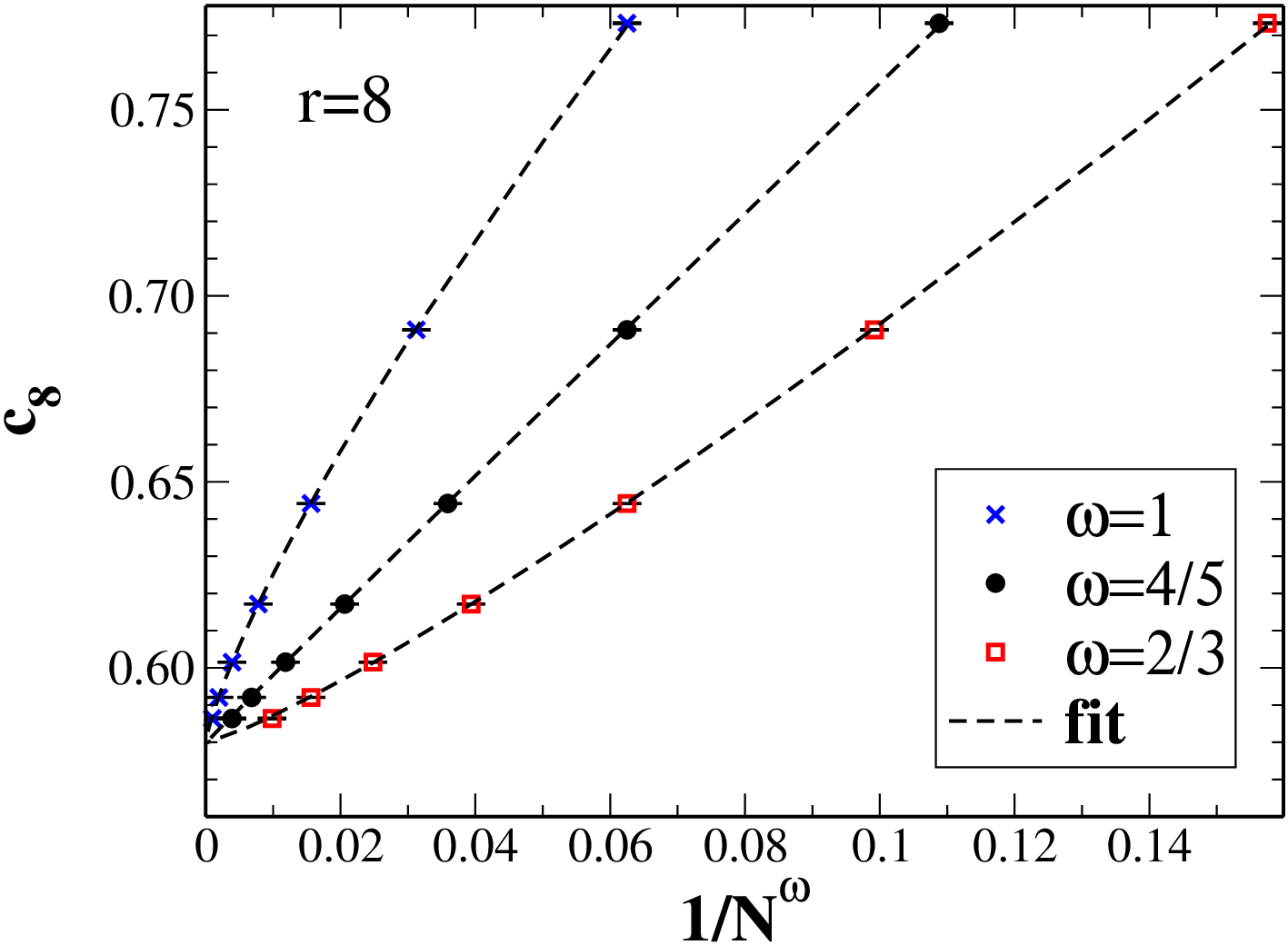}

\includegraphics[clip,scale=0.25]{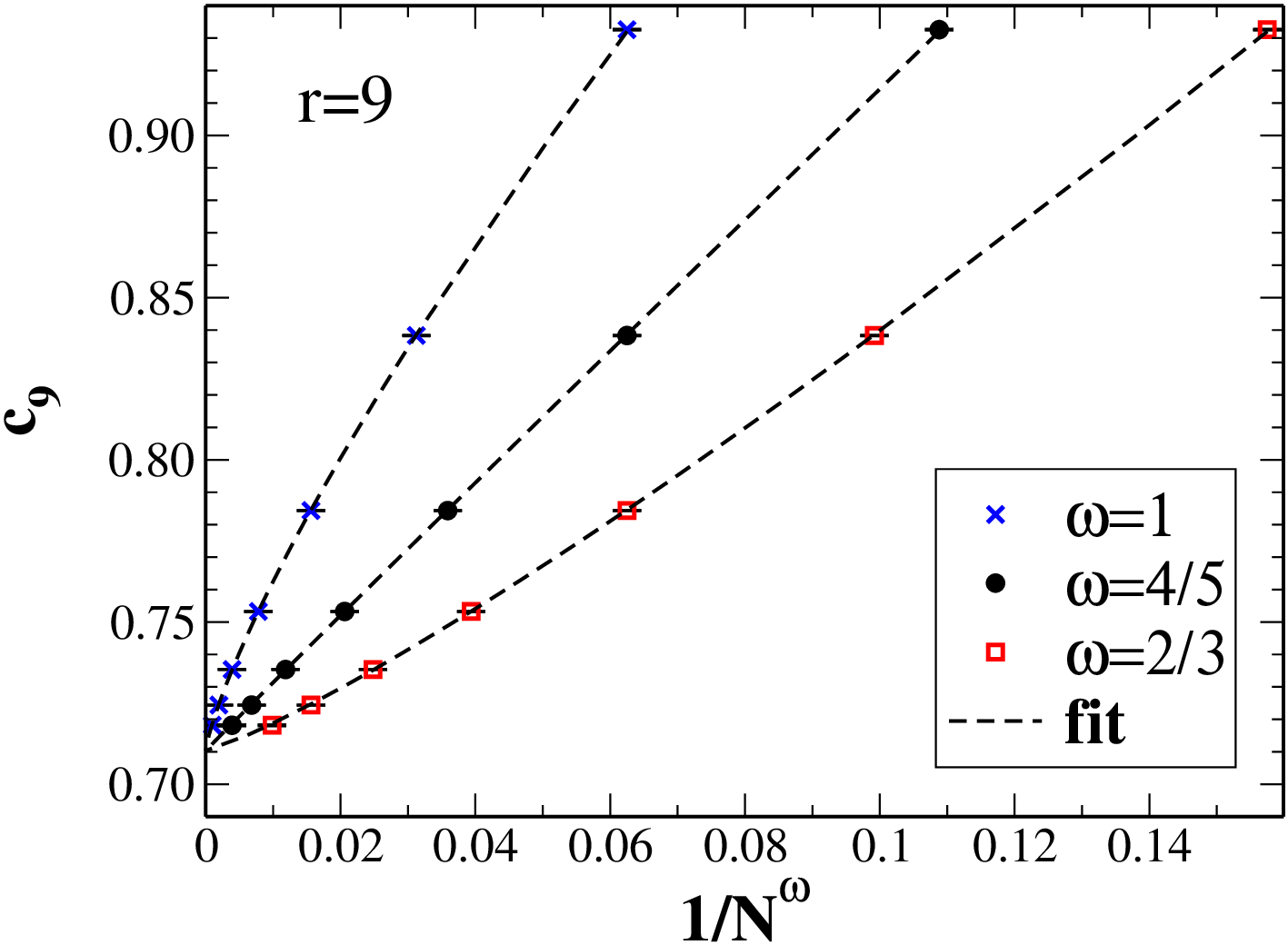}\includegraphics[clip,scale=0.25]{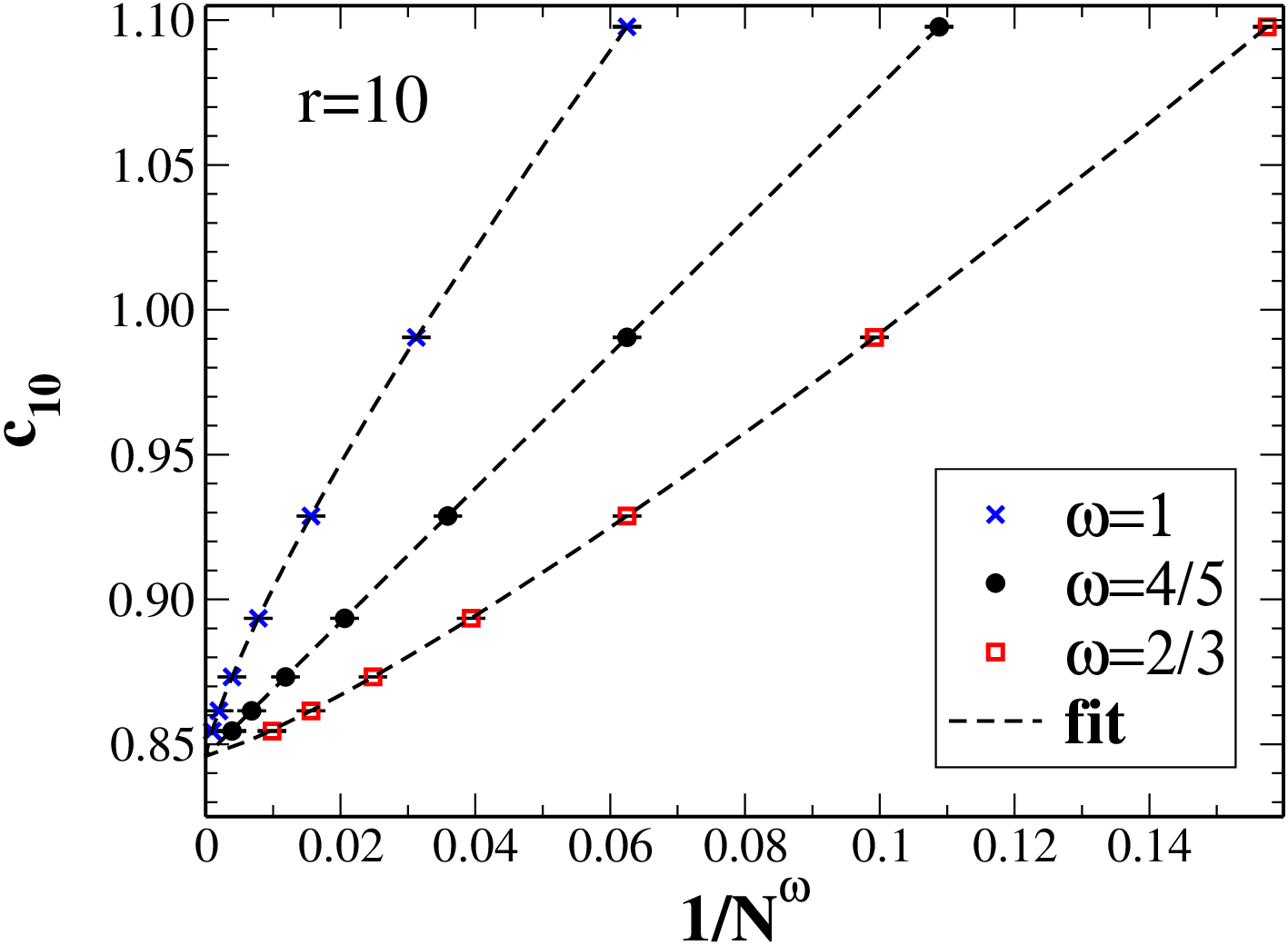}\includegraphics[clip,scale=0.25]{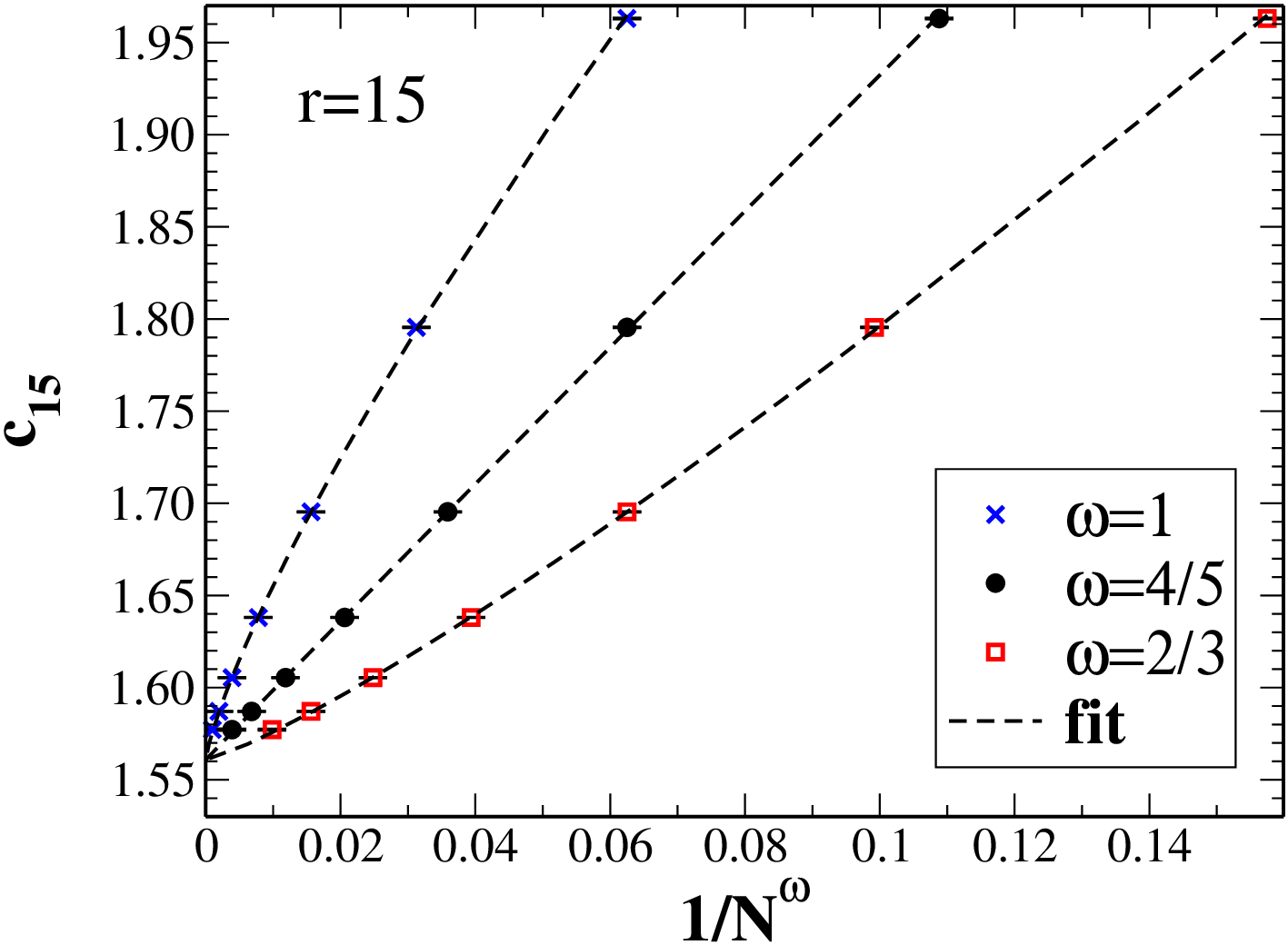}
\caption{\label{fig:Extrapolation}
Extrapolation of the average ground state cost per spin for Bethe
lattices of degree $r=3,4,\ldots,10,$ and 15 on a $1/N^{\omega}$-scale
for $\omega=1$ (${\bf \times}$), $\frac{4}{5}$ (${\bullet}$), and
$\frac{2}{3}$ (${\bf \Box}$).  The intercept of the fit with the
ordinate represents the thermodynamic average (see
Tab.~\ref{alldata}), which varies  weakly with the choice of
$\omega$. For each $r$, a fit to the form $a+b/N^{\omega}$, weighted
by the error, consistently results in $\omega\approx\frac{4}{5}$ (see
Tab.~\ref{power} and Figs.~\ref{fig:Qtest}-\ref{fig:omega}). Only rescaling to
$1/N^{\frac{4}{5}}$ virtually linearizes the extrapolation. 
}
\end{figure*}

\section{$\tau$-EO Algorithm for Bethe Lattices\label{sec:-EO-Algorithm}}
To obtain the numerical results in this paper, we used exactly the
same implementation of $\tau$-EO to find ground states as in
Ref.~\cite{Boettcher03a}, except that we assign to each spin $x_{i}$
in Eq.~(\ref{Heq}) a {}``fitness''
\begin{eqnarray}
\lambda_{i}=\left\lfloor10\frac{x_{i}\sum_{<\dot{,}j>}J_{i,j}x_{j}}{\sum_{<\dot{,}j>}\left|J_{i,j}\right|}\right\rfloor
\end{eqnarray}
Instead of using the local field of each variable, which may vary in
range somewhat between variables due to the fluctuations in the bond
values, we re-scale each variable into the same interval
$[-10,-9,-8,\ldots,+10]$, thereby {}``hashing'' the otherwise
continuous state-space into a discrete set of up to 21 bins. Note that
in this case the sum of the fitnesses is \emph{not} proportional to
the total energy, but good fitness sufficiently correlates with good
costs for local search with EO to succeed, as explained in
Ref.~\cite{Boettcher00}.

\begin{table*}
\caption{\label{alldata} Average cost per spin $\left\langle c_{r}\right\rangle _{N}$,
plotted also in Fig.~\ref{fig:Extrapolation}. The  result for $\langle
c\rangle_{\infty}$ is based on the value for $a$ in Tab.~\ref{power}.}
\begin{tabular}{r|lllllllllll}
\hline 
$N$ & $\left\langle c_{3}\right\rangle _{N}$  & $\left\langle c_{4}\right\rangle _{N}$  & $\left\langle c_{5}\right\rangle _{N}$  & $\left\langle c_{6}\right\rangle _{N}$  & $\left\langle c_{7}\right\rangle _{N}$  & $\left\langle c_{8}\right\rangle _{N}$  & $\left\langle c_{9}\right\rangle _{N}$  & $\left\langle c_{10}\right\rangle _{N}$  & $\left\langle c_{15}\right\rangle _{N}$  &  & \tabularnewline
\hline 
16  & 0.09043(5)  & 0.20163(7)  & 0.3302(1)  & 0.4703(1)  & 0.6182(1)  & 0.7732(2)  & 0.9326(2)  & 1.0976(2)  & 1.9630(3)  &  & \tabularnewline
32  & 0.06934(3)  & 0.16825(5)  & 0.28476(6)  & 0.41254(7)  & 0.54844(8)  & 0.6908(1)  & 0.8383(1)  & 0.9904(1)  & 1.7954(2)  &  & \tabularnewline
64  & 0.05759(2)  & 0.14934(3)  & 0.25882(4)  & 0.37961(4)  & 0.50894(5)  & 0.64416(6)  & 0.78435(6)  & 0.92878(7)  & 1.6953(1)  &  & \tabularnewline
128  & 0.05087(1)  & 0.13842(2)  & 0.24387(2)  & 0.36103(3)  & 0.48600(3)  & 0.61714(4)  & 0.75324(4)  & 0.89347(5)  & 1.63804(6)  &  & \tabularnewline
256  & 0.04717(1)  & 0.13225(1)  & 0.23544(2)  & 0.35003(2)  & 0.47271(2)  & 0.60153(3)  & 0.73536(3)  & 0.87325(3)  & 1.60540(7)  &  & \tabularnewline
512  & 0.04491(1)  & 0.12853(1)  & 0.23026(2)  & 0.34363(2)  & 0.46465(3)  & 0.59203(4)  & 0.72440(4)  & 0.86153(4)  & 1.58710(9)  &  & \tabularnewline
1024  & 0.04338(4)  & 0.12616(3) & 0.22736(7) & 0.33967(8) & 0.4601(1) & 0.5863(2) & 0.7181(2) & 0.8545(2) & 1.5772(4) &  & \tabularnewline
2048  & 0.04302(2)  & 0.12577(4) & 0.22666(7) & 0.3384(1) &  &  &  &  &  &  & \tabularnewline
\hline 
$\infty$  & 0.0420(1)  & 0.1236(1)  & 0.2236(1)  & 0.3352(3)  & 0.4542(5)  & 0.5798(8)  & 0.7103(9)  & 0.8459(9)  & 1.561(1)  &  & \tabularnewline
\hline
\end{tabular}
\end{table*}
\begin{table}
\caption{\label{power} Fit of the data in Tab.~\ref{alldata} to
  $\langle c_r\rangle_N=a+\frac{b}{N^{\omega}}$.}
\begin{tabular}{r|lll|rrc}
\hline
$r$ & $a$ & $b$ &  $\omega$ & ndf & $\chi^2$/ndf & $Q$ \tabularnewline
\hline
3 &  0.041995 &     0.45 &     0.81 & 6 &            92.5 &               0 \tabularnewline
4 &  0.123592 &     0.70 &     0.79 & 6 &            58.6 &               0 \tabularnewline
5 &  0.223676 &     0.97 &     0.80 & 5 &            63.6 &               0 \tabularnewline
6 &  0.335139 &     1.22 &     0.80 & 5 &            30.3 &         $6.3 \times 10^{-31} $\tabularnewline
7 &  0.454128 &     1.45 &     0.79 & 4 &            38.2 &         $5.5 \times 10^{-32}$ \tabularnewline
8 &  0.579745 &     1.72 &     0.79 & 4 &            43.6 &         $1.3 \times 10^{-36}$ \tabularnewline
9 &  0.710283 &     1.97 &     0.79 & 4 &            40.4 &         $6.5 \times 10^{-34}$ \tabularnewline
10 &  0.845875 &     2.32 &     0.80 & 4 &             1.8 &            0.12 \tabularnewline
15 &  1.560930 &     3.65 &     0.79 & 4 &            59.3 &               0 \tabularnewline
\hline
\end{tabular}
\end{table}
\begin{table}
\caption{\label{allfit_log} Fit of the data in Tab.~\ref{alldata} to
  $\langle c_r\rangle_N=a+\frac{b}{N^{\omega}}+\frac{c\ln N}{N}$.}
\begin{tabular}{r|llll|rrc}
\hline
$r$ & $a$ & $b$ &  $\omega$ & $c$ & ndf & $\chi^2$/ndf & $Q$ \tabularnewline
\hline
3 &  0.041654 &     0.61 &     0.78 &     -0.13 & 5 &            37.9 &         5.5e-39 \tabularnewline
4 &  0.123219 &     0.85 &     0.77 &     -0.12 & 5 &            50.1 &               0 \tabularnewline
5 &  0.223430 &     1.06 &     0.79 &     -0.07 & 4 &            75.8 &               0 \tabularnewline
6 &  0.334402 &     1.46 &     0.77 &     -0.20 & 4 &            20.2 &         1.1e-16 \tabularnewline
7 &  0.452726 &     1.86 &     0.76 &     -0.35 & 3 &             7.1 &         8.6e-05 \tabularnewline
8 &  0.577866 &     2.24 &     0.76 &     -0.45 & 3 &             7.6 &         4.3e-05 \tabularnewline
9 &  0.708515 &     2.47 &     0.76 &     -0.42 & 3 &            17.6 &           2e-11 \tabularnewline
10 &  0.845673 &     2.38 &     0.80 &     -0.05 & 3 &             2.1 &             0.1 \tabularnewline
15 &   &    unstable  &      &     &  &             &            \tabularnewline
\hline
\end{tabular}
\end{table}
\begin{table}
\caption{\label{ln23} Fit of the data in Tab.~\ref{alldata} to
  $\langle c_r\rangle_N=a+\frac{b}{N^{\frac{2}{3}}}+\frac{c\ln N}{N}$.}
\begin{tabular}{r|lll|rrc}
\hline
$r$ & $a$ & $b$ &  $c$ &  ndf & $\chi^2$/ndf & $Q$ \tabularnewline
\hline
3 &  0.042773 &    -0.21 &    -0.45 & 6 &          2579.1 &               0 \tabularnewline
4 &  0.125082 &    -0.37 &    -0.74 & 6 &          1792.5 &               0 \tabularnewline
5 &  0.225941 &    -0.57 &    -1.07 & 5 &          1935.1 &               0 \tabularnewline
6 &  0.337383 &    -0.60 &    -1.25 &   5 &          1995.7 &               0 \tabularnewline
7 &  0.456151 &    -0.53 &    -1.36 &  4 &          2392.7 &               0 \tabularnewline
8 &  0.581872 &    -0.61 &    -1.59 &   4 &          2653.7 &               0 \tabularnewline
9 &  0.713004 &    -0.76 &    -1.87 &  4 &          2772.5 &               0 \tabularnewline
10 &  0.851120 &    -1.40 &    -2.60 & 4 &          3212.2 &               0 \tabularnewline
11 &  1.571370 &    -2.70 &    -4.58 & 4 &          3063.2 &               0 \tabularnewline
\hline
\end{tabular}
\end{table}
\begin{table}
\caption{\label{ln45} Fit of the data in Tab.~\ref{alldata} to
  $\langle c_r\rangle_N=a+\frac{b}{N^{\frac{4}{5}}}+\frac{c\ln N}{N}$.}
\begin{tabular}{r|lll|rrc}
$r$ & $a$ & $b$ &  $c$  & ndf & $\chi^2$/ndf & $Q$ \tabularnewline
\hline
3 &  0.041941 &     0.50 &    -0.03 & 6 &            67.9 &               0 \tabularnewline
4 &  0.123668 &     0.70 &    0.01 & 6 &            62.9 &               0 \tabularnewline
5 &  0.223727 &     0.97 &    0.01 & 5 &            64.4 &               0 \tabularnewline
6 &  0.335286 &     1.21 &    0.02 & 5 &            33.7 &         1.4e-34 \tabularnewline
7 &  0.454604 &     1.37 &    0.08 & 4 &            57.3 &               0 \tabularnewline
8 &  0.580229 &     1.66 &    0.07 & 4 &            59.0 &               0 \tabularnewline
9 &  0.710829 &     1.90 &    0.09 & 4 &            54.6 &               0 \tabularnewline
10 &  0.845843 &     2.33 &    -0.01 & 4 &             1.7 &            0.14 \tabularnewline
11 &  1.561210 &     3.44 &    0.16 & 4 &            46.4 &         4.4e-39 \tabularnewline
\hline

\end{tabular}
\end{table}
\begin{table}
\caption{\label{allfit} Fit of the data in Tab.~\ref{alldata} to
  $\langle c_r\rangle_N=a+\frac{b}{N^{\omega}}+\frac{c}{N}$.}
\begin{tabular}{r|llll|rrc}
\hline
$r$ & $a$ & $b$ &  $\omega$ & $c$ & ndf & $\chi^2$/ndf & $Q$ \tabularnewline
\hline
3 &  0.041500 &     0.12 &     0.62 &     0.43 & 5 &            35.8 &           $9 \times 10^{-37}$ \tabularnewline
4 &  0.123099 &     0.32 &     0.68 &     0.49 & 5 &            50.5 &               0 \tabularnewline
5 &  0.223414 &     0.69 &     0.75 &     0.34 & 4 &            76.6 &               0 \tabularnewline
6 &  0.334213 &     0.57 &     0.69 &     0.81 & 4 &            21.4 &         $1.2 \times 10^{-17}$ \tabularnewline
7 &  0.452095 &     0.52 &     0.63 &     1.21 & 3 &             7.0 &        $ 9.9 \times 10^{-5}$ \tabularnewline
8 &  0.576847 &     0.54 &     0.61 &     1.54 & 3 &             5.6 &         0.00073 \tabularnewline
9 &  0.707617 &     0.73 &     0.64 &     1.61 & 3 &            15.0 &         $9.6 \times 10^{-10} $\tabularnewline
10 &  0.845614 &     2.02 &     0.78 &     0.35 & 3 &             2.0 &            0.11 \tabularnewline
15 &   &   unstable &     &    &  &     &     \tabularnewline
\hline
\end{tabular}
\end{table}
\begin{table}
\caption{\label{power23} Fit of the data in Tab.~\ref{alldata} to
  $\langle c_r\rangle_N=a+\frac{b}{N^{\frac{2}{3}}}+\frac{c}{N}$.}
\begin{tabular}{r|lll|rrc}
\hline
$r$ & $a$ & $b$ &  $c$ & ndf & $\chi^2$/ndf & $Q$ \tabularnewline
\hline
3 &  0.041656 &     0.16 &     0.37 & 6 &            34.3 &         $1.2 \times 10^{-41}$ \tabularnewline
4 &  0.123011 &     0.29 &     0.54 & 6 &            42.4 &               0 \tabularnewline
5 &  0.222800 &     0.39 &     0.74 & 5 &            70.5 &               0 \tabularnewline
6 &  0.333995 &     0.50 &     0.91 & 5 &            17.6 &         $1.7 \times 10^{-17}$ \tabularnewline
7 &  0.452686 &     0.64 &     1.02 & 4 &             7.3 &         $7.4 \times 10^{-6}$ \tabularnewline
8 &  0.577985 &     0.75 &     1.22 & 4 &             8.6 &         $6.6 \times 10^{-7}$ \tabularnewline
9 &  0.708256 &     0.87 &     1.41 & 4 &            12.4 &         $4.1 \times 10^{-10}$ \tabularnewline
10 &  0.843511 &     0.93 &     1.74 & 4 &            17.5 &         $2.1 \times 10^{-14}$ \tabularnewline
15 &  1.557000 &     1.55 &     2.63 & 4 &           146.8 &               0 \tabularnewline
\hline
\end{tabular}
\end{table}

To evaluate the proposed $\tau$-EO algorithm, we have benchmarked over
a number of exactly solved instances obtained with a branch-and-bound
method. Such an approach is clearly limited in attainable system
sizes, here $N\leq64$, and can only be executed for a small test-bed
of instances due to the exponential computational cost of exact
methods. For larger systems, we have also applied the $\tau$-EO
algorithm to a small test-bed of 10 instances of size $N=2^{10}$ for
each value of $r$ with 5-times more updates per run and compared
results. In the worst case, about 30\% of the instances for some $r$
showed a systematic error of about 0.1\% or less. Since we sampled
over $n_I\approx10^{3}-10^4$ instances at this $N$, this systematic
error is well below the statistical error of $\sim1/\sqrt{n_I}\gtrsim1\%$.

Alternatively, we can show that averaged properties obtained with EO
are in some sense self-consistent and/or consistent with certain theoretical
predictions. For example, Fig.~\ref{fig:LargeK} shows the
extrapolation for $r\to\infty$ of the already extrapolated
thermodynamic ($N\to\infty$) limit of the average energies for each
$r$ and $N$. Despite of its derivative nature, the EO data still
reproduces the exactly-known ground state energy of SK very
accurately.

Finally, it should be kept in mind that settings which provide
sufficiently accurate averages of a quantity may be less proficient in
determining its higher moments, let alone its entire PDF. Ever higher
moments are dominated by rare events (or the lack thereof for finite
$n_I$) ever deeper in the exponentially suppressed tails of the
PDF. For instance, variances are based on (subtractions involving)
higher moments and thus can be expected to have substantially boosted
systematic errors compared to those quoted for averages.

\section{Numerical Results\label{sec:Numerical-Results}}

We have simulated Bethe lattices with the algorithm discussed in
Sec.~\ref{sec:-EO-Algorithm} for $r$ between 3 and 15, and graph sizes
$n=2^{l}$ for $l=4,5,6,\ldots,11$ to obtain results for ground state
energies. Statistical errors in our averages have been kept small by
generating a large number of instances for each $N$ and $r$, typically
$n_{I}\approx10^{6}$ for $N\leq256$ and $n_{I}\approx10^{3}-10^{5}$
for $N\geq512$.

\subsection{Average Ground-State Properties\label{sub:Average-Ground-State-Properties}}
\begin{figure}
\includegraphics[clip,scale=0.35]{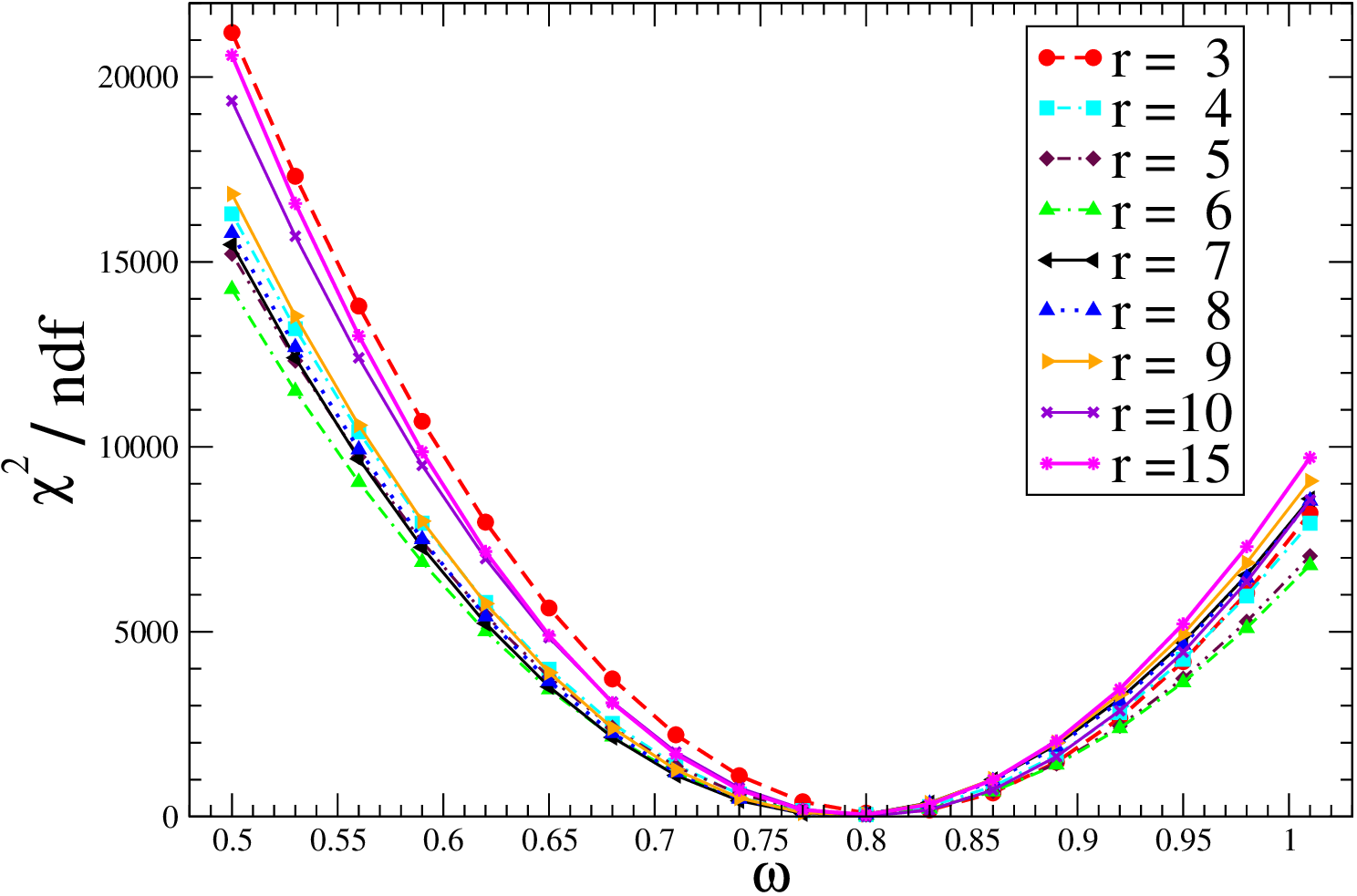}
\caption{\label{fig:Qtest} Plot of the $\chi^2$ per numbers of degrees
  of freedom (ndf) for a fit of the data in Tab.~\ref{alldata} to
  Eq.~(\ref{eq:FSS}) with only the parameters $a$ and $b$ over a range
  of fixed $\omega$, ignoring higher-order corrections
  ($\epsilon\equiv0$). Independently, for each $r$, the data shows a
  distinct minimum near $\omega\approx\frac{4}{5}$.  }
\end{figure}

\begin{figure}
\includegraphics[clip,scale=0.35]{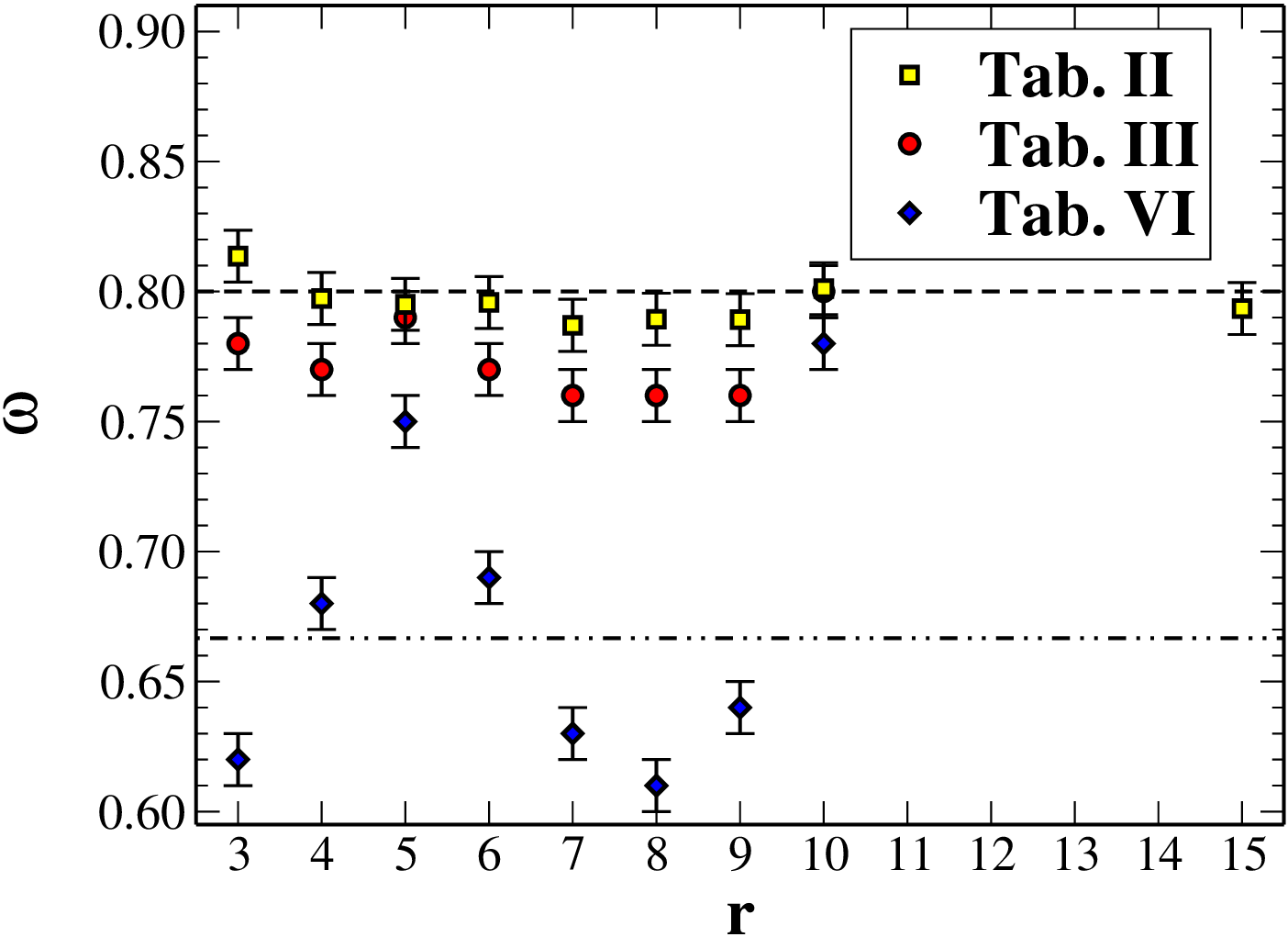}
\caption{\label{fig:omega} Plot of the fitted values for the
  finite-size scaling exponent $\omega$ obtained in Tabs.~\ref{power}
  ($\square$) and~\ref{allfit} ($\circ$) for the available values of
  $r$. Based on an error of $0.01$ estimated from the fits, all values
  from Tab.~\ref{power} are consistent with $\omega=\frac{4}{5}$
  (dashed line), without trend for increasing degree $r$ towards
  $\frac{2}{3}$ (dash-dotted line). The values from Tab.~\ref{allfit}
  behave less uniformly but seem more centered around $\frac{2}{3}$.}
\end{figure}

In Tab.~\ref{alldata}, we list the values of average costs per spin,
$\left\langle c\right\rangle _{N}$. When plotted as a function of
$1/N$ in Fig.~\ref{fig:Extrapolation}, the average costs per spin for
each given $r$ clearly do not extrapolate linearly.  Instead, we
attempt a fit according to Eq.~(\ref{eq:FSS}) with variable
finite-size correction exponent $\omega$, at first ignoring
higher-order corrections ($\epsilon\equiv0$).  Listed in
Tab.~\ref{power}, we find that for the whole range of connectivities
$r$ studied here, the fitted scaling corrections appear to be
consistent with $\omega=4/5$, and a plot of the data in
Fig.~\ref{fig:Extrapolation} on a rescaled abscissa, $1/N^{4/5}$,
produces linear scaling. In
Fig.~\ref{fig:Qtest}, we assess the quality of a fit, free of any
assumptions about unknown higher-order corrections, by fixing the value
of $\omega$ over a range and then fitting for the remaining two
parameters, $a$ and $b$, in Eq.~(\ref{eq:FSS}). Each of the
\emph{independent} data sets for $r=3,\ldots,10,15$ exhibits a strong
preference for $\omega\approx4/5$. In Fig.~\ref{fig:omega}, we plot
those fitted values for $\omega$, which within the range of degree
values $r$ studied here suggest no trend toward the value of
$\omega_{SK}=\frac{2}{3}$ expected for the SK-limit $r\to\infty$,
irrespective of bond
distribution~\cite{Bouchaud03,Boettcher05e,Aspelmeier07,Palassini08}. Plotting
the data for $\omega=\frac{2}{3}$ in Fig.~\ref{fig:Extrapolation}
clearly does not produce a linear extrapolation as in
Refs.~\cite{Boettcher03a,Boettcher03b} for discrete bonds.

In Tabs.~\ref{allfit_log}-\ref{power23}, we have considered
alternative fits to the data involving also higher-order corrections
$\epsilon(N)$ to the finite-size corrections, using the full form of
Eq.~(\ref{eq:FSS}). While there are some theoretical
results~\cite{parisi:93,parisi:93b}, obtained for the internal energy
near $T_c$ in SK, possibly justifying $\omega=\frac{2}{3}$ also below
$T_c$, little is know to higher order. The expansion for the free
energy at $T_c$ in Refs.~\cite{parisi:93,parisi:93b} provided
corrections of the form $c\ln(N)/N$, which could be argued to
eventually affect also the ground state energy. We have therefore
attempted to fit the data also with higher-order corrections of that
form, see Tab.~\ref{allfit_log}, and simply $c/N$, another plausible
form. [Unfortunately, a five-parameter fit with a higher-order
  correction of $c/N^\alpha$ does not provide stable results for our
  data.]  Assuming either correction provides the best-quality fit to
the remaining four parameters, with the lowest $\chi^2$ value relative
to the remaining numbers of degree of freedom (ndf), shown in
Tabs.~\ref{allfit_log} and~\ref{allfit}. The former fit yields almost
identical results to that without higher-order correction,
Tab.~\ref{power}, only that the values for $\omega$ are consistently
shifted down by a small amount, see Fig.~\ref{fig:omega}. The fit with
$1/N$ drastically changes the fitted values. The values for $\omega$
are now somewhat consistent with $\frac{2}{3}$, but vary widely with
degree $r$ in Fig.~\ref{fig:omega}. Fixing $\omega=\frac{2}{3}$
improves the quality of fit (with one less parameter) for $1/N$
corrections to the best fit overall, see Tab.~\ref{power23}. But it
becomes entirely inconsistent with $\ln(N)/N$ corrections, as shown in
Tab.~\ref{ln23}, unlike for fixed $\omega=\frac{4}{5}$ in
Tab.~\ref{ln45}.

\begin{figure}
\includegraphics[clip,scale=0.33]{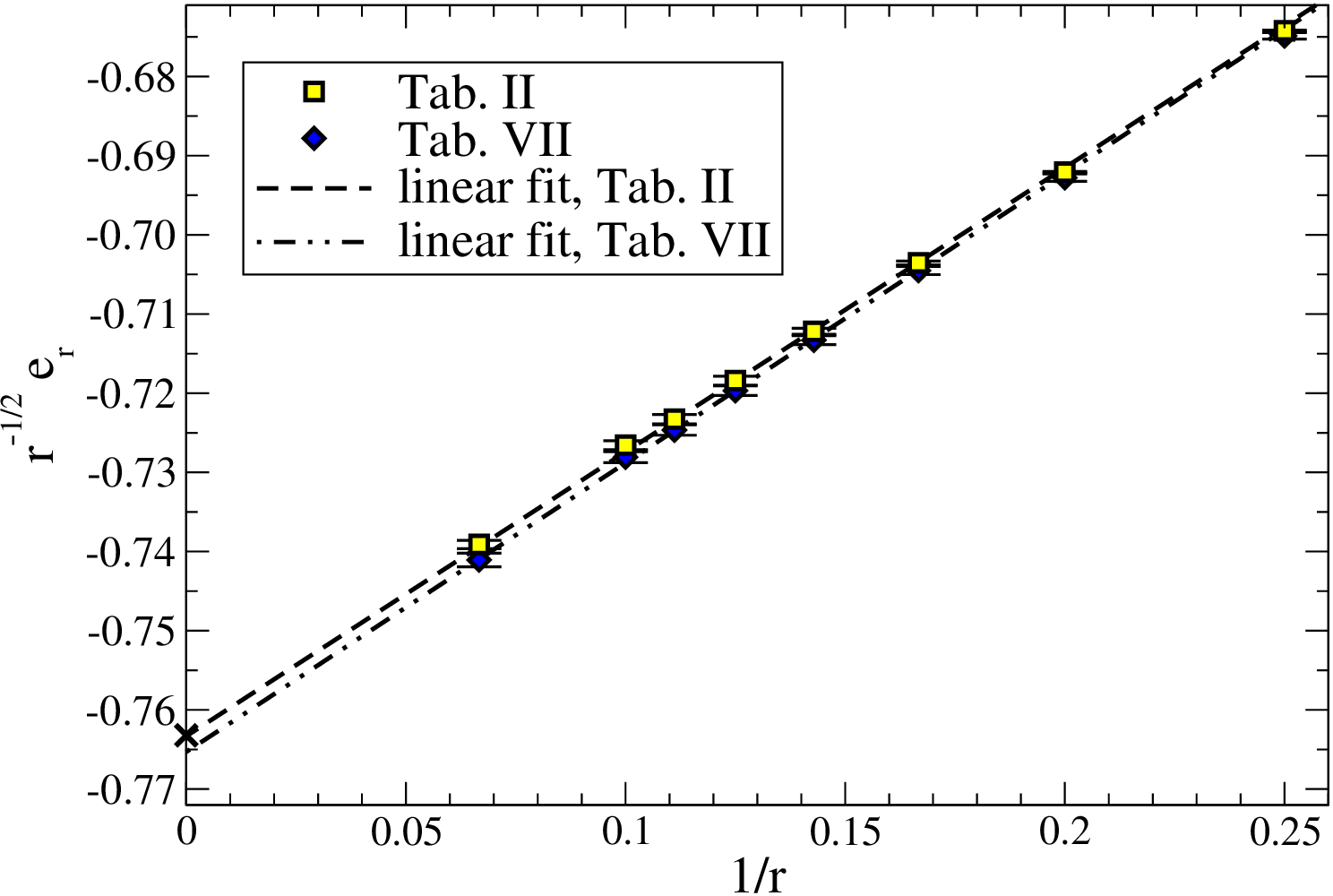}
\caption{\label{fig:LargeK} Extrapolated values of the energy
  densities $\left\langle e_{r}\right\rangle _{\infty}$ obtained from
  the cost densities $\left\langle c_{r}\right\rangle _{\infty}$
  listed in Tab.~\ref{alldata} (yellow $\square$). The data is plotted
  as a function of inverse degree, $1/r$, and rescaled by a root of
  the degree, such that the large-$r$ limit approaches the SK
  model. Similar to the corresponding plot in
  Ref.~\cite{Boettcher03a,Boettcher03b}, the data appears to reach the
  SK limit virtually on a linear trajectory: a linear fit (dashed
  line) to all data projects the ground-state of the SK model,
  $\left\langle e_{SK}\right\rangle =-0.76317$ (marked by $\times$) as
  $-0.76332$, or to within an error of $<0.1\%$. An extrapolation of
  corresponding data from Tab.~\ref{power23} (blue $\diamond$) misses
  the SK value noticeably (dash-dotted line).  }
\end{figure}

In should be remarked that in all cases, the confidence-of-fit $Q$ is
essentially zero in light of the rather tight error bars obtained from
the statistics for the average cost densities. We would argue that
this is due to the inherent limitations of fitting this data down to
small system size $N$ with an asymptotic form valid for large $N$,
Eq.~(\ref{eq:FSS}), as a stand-in for an entirely unknown function of
$N$. Hence, these fits should not be dismissed on the basis of $Q$
alone.

The extrapolated values for $a=\left\langle c_{r}\right\rangle
_{\infty}$ obtained in Tab.~\ref{power} are also listed in
Tab.~\ref{alldata}. To demonstrate the quality of the extrapolation,
we plot in Fig.~\ref{fig:LargeK} the derived values for the energy
densities by way of Eq.~(\ref{eq:averageE}), $\left\langle
e_{r}\right\rangle _{\infty}=2\left\langle c_{r}\right\rangle
_{\infty}-\left\langle b_{r}\right\rangle _{\infty}$, using
Eq.~(\ref{eq:averageB}). As in Refs.~\cite{Boettcher03b,Boettcher03a},
already a linear fit to the extrapolated values reproduces the exactly
known value of the SK model~\cite{MPV,crisanti:02,Oppermann07} to
within an error of $<0.1\%$, confirming to a high degree the numerical
accuracy of the data. The obtained slope of the extrapolation, called
$f_{1}$ in Ref.~\cite{Tria02}, which is not expected to be a universal
quantity, evaluates to $f_{1}\approx0.36$, much larger than for the
corresponding problem with $\pm J$
bonds~\cite{Boettcher03a,Boettcher03b}. The extrapolated costs in
Tab.~\ref{power23}, also plotted as energies in
Fig.~\ref{fig:Extrapolation}, markably miss  the SK value.

\begin{figure}
\includegraphics[clip,scale=0.35]{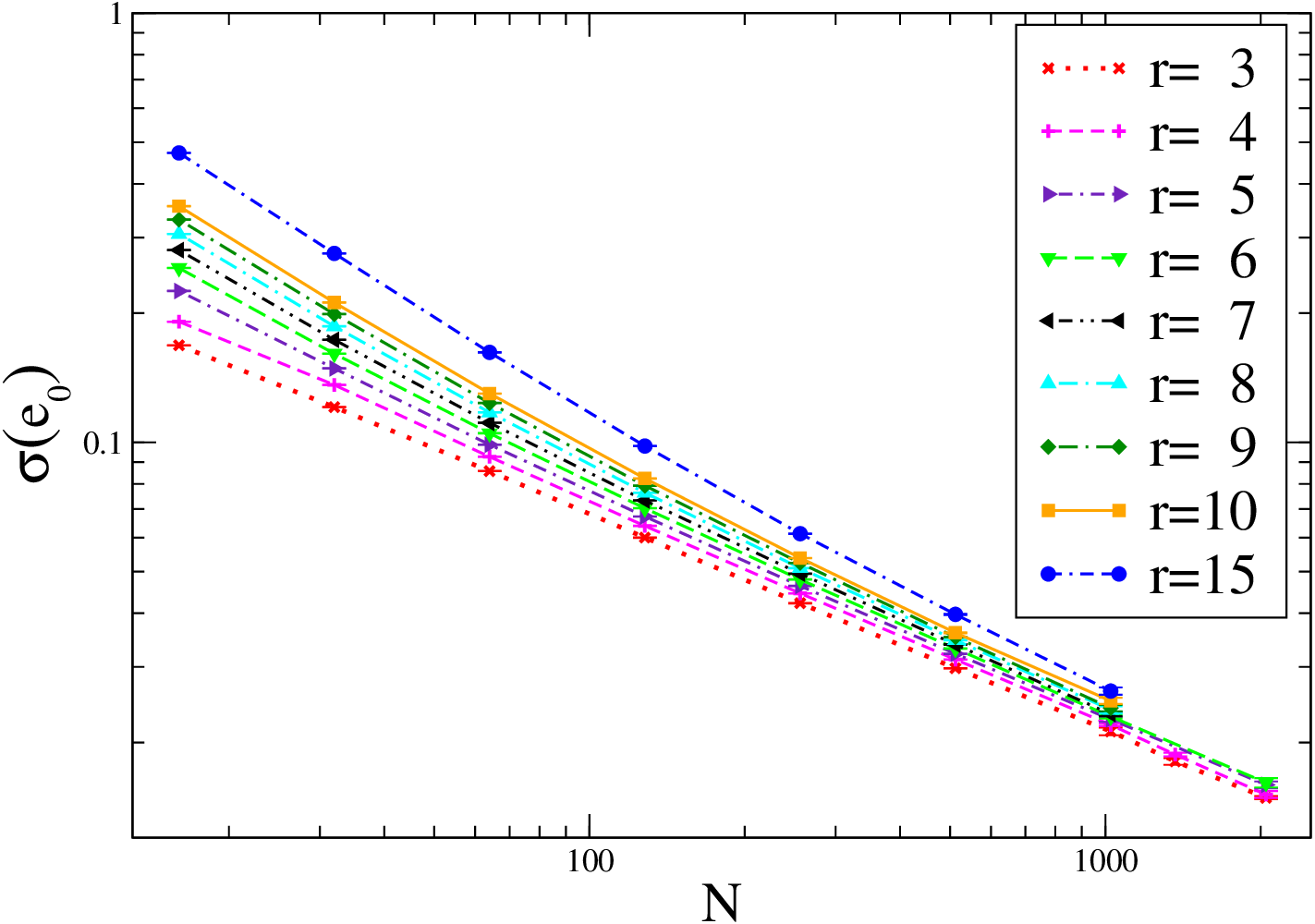}
\includegraphics[clip,scale=0.35]{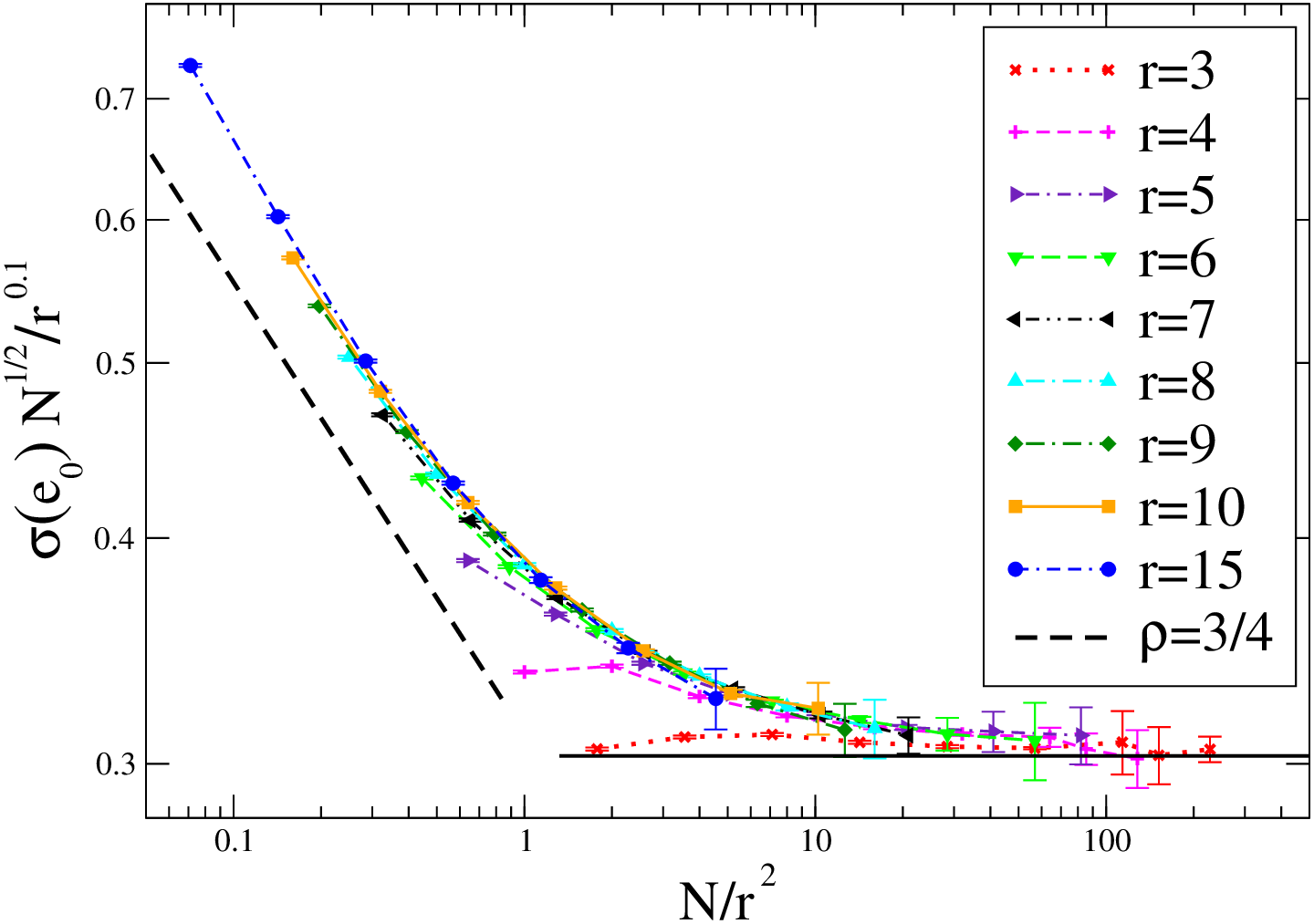}
\caption{\label{fig:Scaling-collapse-for GSEF}
Scaling collapse for the deviation in the ground state fluctuations
for the energy densities, $\sigma_{N}(e)$.  The raw data for the Bethe
lattices is plotted for the available degrees $r$ on top. On the
bottom, the same data is plotted, now rescaled by the indicated powers
of the degree, with the deviation also multiplied by $\sqrt{N}$, such
that trivial, normal fluctuations should plateau (as indicated by the
continuous horizontal line). For fixed degree, systematic deviations
appear only for sufficiently small $N\lesssim r^{2}$, consistent with
a non-trivial scaling $\sigma(e)\sim N^{-\rho}$ at a value of
$\rho=\frac{3}{4}$ (dashed line) or even $\rho=\frac{5}{6}$, as
expected for the SK model. If taken at face-value, one would conclude
that for any fixed degree at large $N$ merely trivial scaling is
obtained and only the SK model (or any model with degree growing as
$r\gg\sqrt{N}$) has anomalous scaling, $\rho>\frac{1}{2}$.}
\end{figure}

\begin{figure}
\includegraphics[clip,scale=0.35]{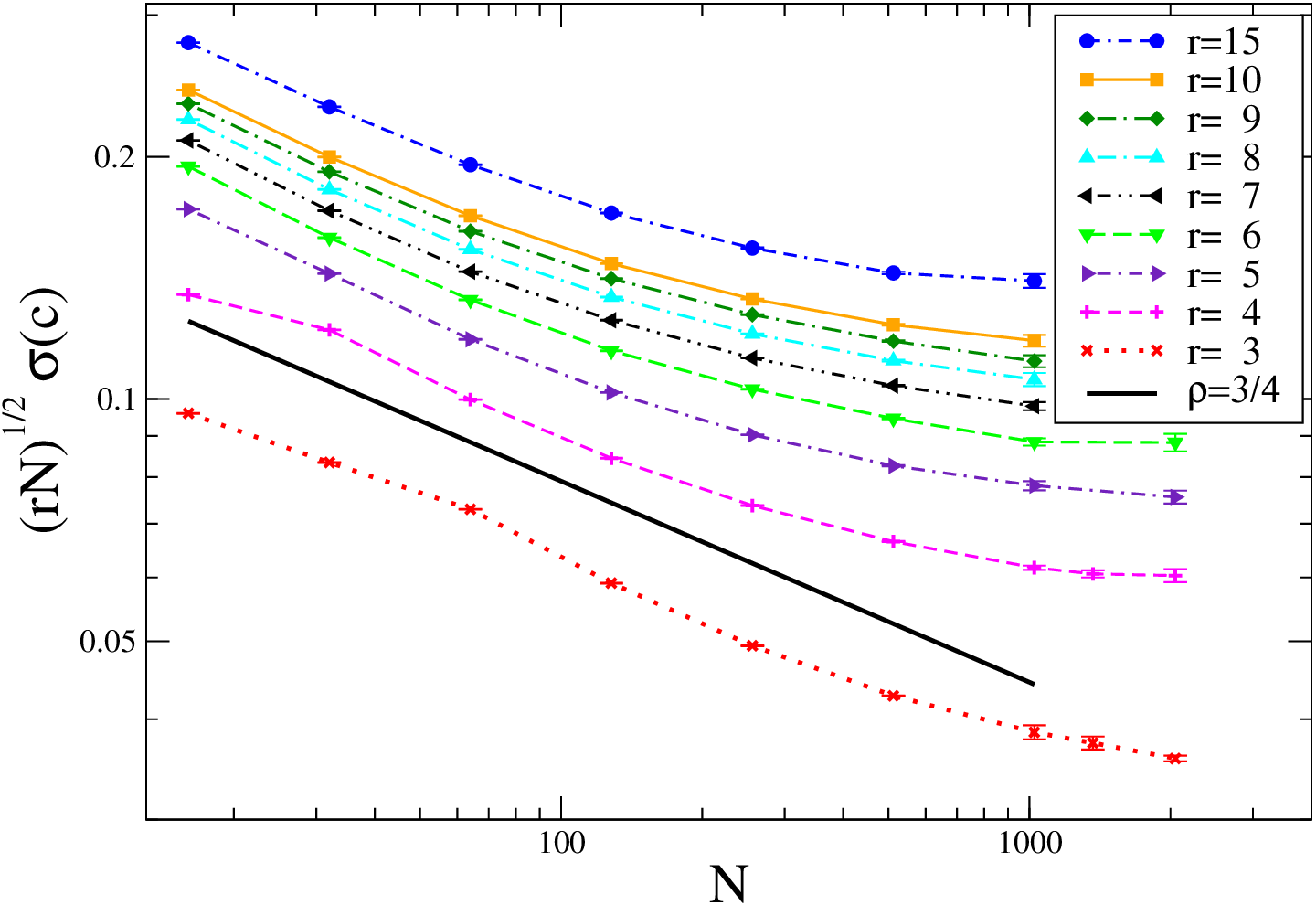}

\caption{\label{fig:Scaling-collapse-for GSCF}
Plot for the deviation in the ground state fluctuations for the cost
densities, $\sigma_{N}(c)$.  The data for the Bethe lattices is
plotted for the available degrees, $r=z$, with the deviation
multiplied by $\sqrt{N/r}$, such that trivial, normal fluctuations
should plateau. No collapsible scaling regimes emerge, such as a
non-trivial scaling $\sigma(e)\sim N^{-\rho}$ with $\rho=\frac{3}{4}$
(continuous line) and much less $\rho=\frac{5}{6}$, as expected for
the SK model. Instead, all data appears to steadily approach trivial
scaling at much larger system sizes $N$.}
\end{figure}

\subsection{Ground-State Energy and Cost Fluctuations\label{sub:Ground-State-Energy-Fluctuations} }
While the previous section has demonstrated the numerical accuracy of
the data at the level of the average of observables, higher moments or
a measure of the entire probability density function (PDF) of ground
state energy fluctuations provide a more confusing picture. No clear
conclusion can be reach on the basis of this data, even when heeding
the implications of Sec. \ref{sec:Sampling-Ground-States}. In light of
that discussion, we first present the (as we believe, incorrect)
extrapolation for the ground-state \emph{energy} fluctuations,
followed by the corresponding discussion for the cost fluctuations.

In Fig.~\ref{fig:Scaling-collapse-for GSEF}, we show the raw data for
the deviations of the energy densities $\sigma_{N}\left(e\right)$ and
their apparent collapse. All the data seems to approach trivial,
normal fluctuations for sufficiently large system sizes, with a
cross-over at ever higher degree $r$. Similar normal fluctuations for
this model have been claimed by Ref.~\cite{Bouchaud03}. In fact,
rescaling the data for a collapse indicates that the cross-over
between system size and degree in the Bethe lattice occurs at $N\sim
r^{2}$. Taken at face value, this would be a remarkable result. While
at any fixed, finite degree a trivial scaling is reached, \emph{only}
in the SK-limit, where system size and degree would scale direct
proportionally, $N\sim r$, we would flow towards the left onto the
non-trivial branch of the scaling in
Fig.~\ref{fig:Scaling-collapse-for GSEF}. Hence, the Bethe lattice
results for Gaussian bonds (unlike for $\pm J$ bonds, as we will show
in Ref.~\cite{Bo_unpub}) are disconnected from the SK-limit.  Although
this may also explain the unusual finite-size scaling corrections
$\omega\approx0.8$, as well disconnected from the SK-limit, we believe
that this data collapse does not probe the true disorder-induced
frustration.  As we have shown in
Sec.~\ref{sec:Sampling-Ground-States}, energy fluctuations are
dominated by the variance in the Gaussian distribution of $N$ bonds
(although ever less so for larger $r$).

Remarkably, when we plot the deviations $\sigma_{N}\left(c\right)$ for
the cost density in Fig.~\ref{fig:Scaling-collapse-for GSCF}, a far
more difficult-to-interpret picture results. Unlike for the energy
densities in Fig.~\ref{fig:Scaling-collapse-for GSEF}, there is no
apparent cross-over but the data instead veers steadily towards normal
fluctuations for much larger system sizes. While the data does not
show any noticeable statistical errors, it is impossible to exclude a
systematic bias in the sampling of ground states with the heuristic,
that could manifest itself in a smooth drift away from any potential
non-trivial scaling. We can only eliminate the systematic error up to
$N\leq64$ through comparisons with exact ground states obtained with a
branch-and-bound algorithm. Above such sizes, we can only argue for
small systematic errors based on the internal consistency of the
average cost or energies, as is displayed, e.~g., in
Figs.~\ref{fig:Extrapolation} and~\ref{fig:LargeK}. Though, this may
prove insufficient to guarantee similar fidelity for higher cumulants,
like the deviations in the cost. But if we assume sufficient accuracy
for this data, it would imply that even for the cost deviations, like
for the energies before in Fig.~\ref{fig:Scaling-collapse-for GSEF},
ultimately pure normal fluctuations may result either way when
Gaussian bonds are considered. It would not be unusual to find
extended transient behavior in spin glasses with Gaussian
bonds~\cite{BoCo}.

\section{Conclusion\label{sec:Conclusion}}
We have found surprising differences in the finite-size scaling
behavior between a continuous, Gaussian bond distribution and previous
results for a bimodal, $\pm J$
distribution~\cite{Boettcher03a,Boettcher03b} for spin glasses on
Bethe lattices of degree $r$. While either distribution leads to
equivalent results for thermodynamic averages that smoothly
extrapolate to the exactly known SK results with identical scaling in
$r$ for $r\to\infty$, the finite-size corrections term $b/N^{\omega}$
not only differ in the correction amplitude $b$ but possibly in the
scaling exponent $\omega$ itself. Only when higher-order corrections
are postulated, more consistency can be obtained with the value
$\omega\approx\frac{2}{3}$ found for discrete bonds on Bethe lattices
and for SK with either bond distribution. The value obtained here for
Gaussian bonds, $\omega\approx\frac{4}{5}$, raises the question about
an eventual cross-over to the SK value at higher $r$. No tendency
toward such a cross-over is apparent in our study up to $r=15$. In
light of that, the vague expectation of some uncooperative
higher-order corrections to $\omega=\frac{2}{3}$ seems preferable, but
one might be forgiven to be struck by the solid persistence of
$\omega=\frac{4}{5}$ suggested by Figs.~\ref{fig:Extrapolation}
and~\ref{fig:omega}.

Our study of fluctuations in the ground-state properties is not
successful in determining clearly the scaling behavior of the
deviations. But it sends a cautionary note about the origin of
fluctuations and the interpretation of data when simulating spin
glasses on sparse graphs with continuous  bonds or a randomly
fluctuating geometry.

\section*{Acknowledgments}
This work has been supported by the U.~S. National Science Foundation
through grant  DMR-0812204.

\bibliography{/Users/stb/Boettcher}
\end{document}